\documentclass[%
reprint,
amsmath,amssymb,
aps,
]{revtex4-1}

\pdfoutput=1

\usepackage{mathtools}
\usepackage{tikz}
\usetikzlibrary{matrix,arrows,decorations.pathmorphing,decorations.pathreplacing}
\newcommand{\EQ}[1]{\begin{equation}\begin{split} #1
\end{split}\end{equation}}

\def\Bk{{\boldsymbol k}}
\def\BX{{\boldsymbol x}}
\def\BY{{\boldsymbol y}}
\def\BP{{\boldsymbol p}}
\def\BQ{{\boldsymbol q}}

\usepackage{amsmath}
\usepackage{slashed}
\usepackage{amssymb}
\usepackage{braket}
\usepackage{graphicx}
\usepackage{amsmath}
\usepackage{braket}
\usepackage{epigraph}
\usepackage{tensor}
\usepackage{enumerate}
\usepackage{subfig}
\usepackage{verbatim}
\usepackage{tikz}
\usetikzlibrary{matrix,arrows,decorations.pathmorphing,decorations.pathreplacing}
\usepackage{amsmath,amssymb,euscript,array,mathrsfs,appendix,ctable,marvosym}
\usepackage{arydshln}
\usepackage{todonotes}
\usepackage{graphicx}
\usepackage[normalem]{ulem}

\usepackage{tikz}
\usepackage{pgfplots}
\pgfplotsset{/pgf/number format/use comma,compat=newest}
\usepackage{color}

\usepackage{hyperref}

\usepackage{braket}
\usepackage{amsfonts}
\usepackage{epstopdf}
\usepackage{color}
\usepackage{graphicx}
\usepackage{dcolumn}
\usepackage{bm}
\usepackage{tensor}

\usepackage{braket}

\begin{document}


\title{Decoherence, discord and the quantum master equation for cosmological perturbations}

\author{Timothy J.~Hollowood}
\email{t.hollowood@swansea.ac.uk}

\author{Jamie~I.~McDonald}
\email{pymcdonald@swansea.ac.uk}
\affiliation{%
Department of Physics, Swansea University, Singleton Park, Swansea, SA2 8PP, UK}%


%

\date{\today}

\begin{abstract}
We examine environmental decoherence of cosmological perturbations in order to study the quantum-to-classical transition and the impact of noise on entanglement during inflation. Given an explicit interaction between the system and environment, we derive a quantum master equation for the reduced density matrix of perturbations, drawing parallels with quantum Brownian motion, where we see the emergence of fluctuation and dissipation terms. Although the master equation is not in Lindblad form, we see how typical solutions exhibit positivity on super-horizon scales, leading to a physically meaningful density matrix. This allows us to write down a Langevin equation with stochastic noise for the classical trajectories which emerge from the quantum system on super-horizon scales. In particular, we find that environmental decoherence increases in strength as modes exit the horizon, with the growth driven essentially by white noise coming from local contribution to environmental correlations. Finally, we use our master equation to quantify the strength of quantum correlations as captured by discord. We show that environmental interactions have a tendency to decrease the size of the discord and that these effects are determined by the relative strength of the expansion rate and interaction rate of the environment. We interpret this in terms of the competing effects of particle creation versus environmental fluctuations, which tend to increase and decrease the discord respectively.
\end{abstract}

\maketitle


\section{\label{sec:level1}Introduction}

The inflationary paradigm \cite{G1,G2} successfully describes the present-day uniformity of the Universe on large scales, as well as accounting for small fluctuations in the cosmic microwave background (CMB) \cite{P1,P2,P3,P4}, leading to a nearly-scale-invariant power spectrum of temperature anisotropies. According to the inflationary hypothesis, the origin of these inhomogeneities can be traced to primordial quantum fluctuations produced during the inflationary epoch, which become stretched and amplified across the sky by the quasi-de Sitter expansion. In this sense, primordial fluctuations sow the seeds for the large-scale structure observed today. However, both large-scale structure and even the CMB are treated in an essentially classical way, in that they are described by classical probabilities. A question of fundamental importance is therefore to understand how, and in what sense these fluctuations are rendered classical as they journey outside the horizon.

For free fields in (quasi-)de Sitter there is a standard answer to this question. Since the fluctuations are isolated, they can be described by a pure state whose corresponding wavefunction
\begin{equation}\label{wavefunction}
\Psi[\chi_\Bk] =\left( \frac{2 \text{Re}\,\Omega_{\Bk} }{\pi} \right)^{1/2} \exp\left( - \Omega_{\Bk} |\chi_\Bk|^2 \right),
\end{equation}
becomes highly squeezed (in the momentum direction) as modes exit the horizon. As explained in \cite{KP,Kiefer:2008ku,Martin:2012pea} expectation values calculated in this squeezed state can be described as averages over a set of classical stochastic trajectories in analogy with a Gaussian wavepacket of a free-particle. Nevertheless, the wave function is still a pure state with long range coherence in field space and the interpretation in terms of classical stochastic trajectories is seen to be special to a free theory. It is interesting to go beyond the free theory and study the quantum-to-classical transition when there are non trivial interactions.

Indeed, in reality, fluctuations are not truly isolated and will interact with other fields (at the very least they must interact with metric fluctuations) playing the role of an ``environment". Therefore, one necessarily has to understand the emergence of classicality in \textit{open} systems, and in what sense a classical stochastic description emerges. This is the subject of open quantum systems \cite{BP}, which provides a formalism to analyse the reduced density matrix of the particular system (and its evolution via a quantum master equation) by tracing out the environmental degrees of freedom. In particular, this formalism allows one to study the process of environmental decoherence, which at the simplest level, leads to the decay of quantum interference in the system. It also allows one to understand how, starting from an initially quantum state, classical \textit{stochastic} dynamics emerges due to interaction \textit{with an environment} -- this is of course more challenging than the situation in the free theory where the classical description is rather trivial.

Steps towards understanding the evolution of the density matrix during inflation in the presence of an external environment have already been made in refs.~\cite{KP,KPS2000,Kiefer:2006je} where a phenomenologically motivated ansatz of the form
\begin{align}\label{ansatz}
&\bra{\chi_{\Bk}} \hat{\rho} \ket{\tilde{\chi}_{\Bk}} \equiv \rho(\chi_{\Bk}, \tilde{\chi}_{\Bk}) = \left( \frac{2 \text{Re}\,\Omega_{\Bk} }{\pi} \right) \nonumber \\
&\times \exp\big[ - \Omega_{\Bk}(\eta) |\chi_\Bk|^2 - \Omega^*_{\Bk}(\eta) |\tilde{\chi}_\Bk|^2 - \frac{\xi}{2}\left| \chi - \tilde{\chi}\right|^2 \big]\ ,
\end{align}
was considered, where $\Omega_\Bk$ are simply the free-solutions associated to eq.~(\ref{wavefunction}). This ansatz consists essentially of adding by hand an additional factor $\exp[-\xi/2 \left| \chi - \tilde{\chi}'\right|^2]$ to the free density matrix associated to (\ref{wavefunction}), where $\xi$ is assumed to be a constant determined by the details of the environment. Plainly eq.~(\ref{ansatz}) is heuristic and warrants further investigation in a concrete model. For instance, $\Omega_\Bk(\eta)$ will obviously receive corrections of order the coupling strength with the environment, thereby incorporating new features into the power spectrum (which is proportional to $[\text{Re}\,\Omega_\Bk]^{-1}$). Similarly, one would expect, and as we shall see, that $\xi$ has some time and momentum dependence $\xi_\Bk(\eta)$. What is needed is to consider particular interactions in order to formulate a more comprehensive analysis of the evolution of the density matrix allowing one to build on the work of, e.g., \cite{KPS1998,BLM} which sketched the implications of particular environment-system couplings. In particular one would like to examine the time-evolution of $\Omega_{\Bk}(\eta)$ and particularly $\xi_\Bk(\eta)$ in a specific model coupling fluctuations to another field.

Under favourable conditions, one can then obtain a quantum master equation for the time-evolution of $\rho$ into which the ansatz (\ref{ansatz}) can be substituted to determine the evolution equations for $\Omega_{\Bk}(\eta)$ and $\xi_\Bk(\eta)$.

In order to have a model of environmental interactions which is amenable to explicit calculation, it is very useful to have one that maintains the description in terms of individual momentum modes of the perturbations $\chi_\Bk$. This leads us to consider the model developed by Boyanovksy \cite{Boy1, Boyanovsky:2015jen} where the interaction is between the ``system" field $\phi$ and another ``environment field" $\varphi$ coupled via
\begin{equation}\label{interaction}
S_\text{int} = \int d^4 x\, \sqrt{-g} \lambda \phi \varphi^2 = \int d^4 x\, a(\eta) \lambda \, \chi \psi^2,
\end{equation}
where $\chi = a \phi$ and $\psi = a \varphi$ are the usual conformally rescaled fields. Note that the special form of the coupling here, linear in $\chi$ (i.e.~$\phi$), means that the interaction does not mix the different momentum modes $\chi_\Bk$ up to and including $\mathcal{O}(\lambda^2)$, and so at lowest order we can still consider the problem mode-by-mode providing a great computational simplification.
In \cite{Boyanovsky:2015jen}, Boyanovksy was focussed more on using the master equation to derive corrections to the power spectrum rather than the details of decoherence and the quantum-to-classical transition. In this paper, we shall develop his analysis and use the interaction (\ref{interaction}) as a toy-model in which to address the evolution of decoherence and the onset of classical stochastic behaviour as modes exit the horizon.

There are different formalisms for modelling the effects of interactions with an environment involving apparently rather different approximations. On the one hand, via a master equation for the reduced density matrix, and on the other, via the {\it influence action\/}. The latter approach can be shown, under suitable circumstances, to be equivalent to the master equation approach in flat space problems \cite{Boy1}, but it is by no means clear whether this will be true in the inflationary context. Although it seems the key observable of the fluctuations -- the power spectrum -- can be calculated in either approach with agreement \cite{Boyanovsky:2015jen}. Some previous works on decoherence in the inflationary context have concentrated on the influence action formalism \cite{Sakagami:1987mp,Lombardo:2005iz}; however, the master equation route offers the advantage of allowing one to follow the quantum state explicitly as expansion and decoherence proceeds and provides a way to calculate the growth of entropy and assess the degree to which the perturbations are classical. Other important work on cosmological perturbations, decoherence and the quantum-to-classical transition includes \cite{Albrecht:1992kf,Polarski:1995jg,Lesgourgues:1996jc,Kiefer:1998qe,Kiefer:1998pb,Campo:2008ij,Kiefer:2006je,Martineau:2006ki,Burgess:2006jn,Kiefer:2008ku,Martin:2012pea,Burgess:2014eoa,Burgess:2015ajz,Matacz:1996gk,EN}.
\\
\\
\noindent \textbf{Key results}
\\
\\
\noindent 1. By deriving and solving evolution equations for $\xi_{\Bk}(\eta)$ and $\Omega_{\Bk}(\eta)$ we are able to show that $\Omega_{\Bk}(\eta)$ does indeed receive additional $\lambda$-dependent corrections. For instance we find that
\begin{align}
&\text{Re}\left[\Omega_{\Bk}(\eta) \right] \nonumber \\
&= k^3 \eta^2 \exp \left[ \frac{\lambda^2}{12 \pi^2 H^2} \Big( \log ^2 \frac{\eta}{\eta_0} - \frac{4}{3} \log \frac{\eta}{\eta_0}\Big)\right]
+O\left( \eta^3\right).
\end{align}
Therefore by direct computation of the density matrix we are able to reproduce, via an alternate calculational route, the resummed corrections to the power spectrum (which is proportional to $1/\text{Re} [\Omega_{\Bk}]$) found by Boyanovksy \cite{Boyanovsky:2015jen}. On super-horizon scales these corrections are dominated by non-local correlations of the ``environment field" $\varphi$, corresponding to long-time correlations, \textit{i.e.} memory effects, as noted in \cite{Boyanovsky:2015jen}.
\\
\\
2. We find that decoherence, as characterised by $\xi_\Bk(\eta)$, exhibits time-dependence, oscillating on sub-horizon scales ($k \eta \gg 1$) and growing monotonically outside the horizon ($k\eta \ll 1$) -- a manifestation of the fact that by virtue of the operator being relevant the factor $a(\eta)$ in the RHS of (\ref{interaction}) leads essentially to an interaction strength which grows in time. By contrast to $\Omega_\Bk(\eta)$, we find that the super-horizon evolution of the decoherence parameter, $\xi_\Bk(\eta)$ is driven by the \textit{local} environment correlators, in other words, by white noise relating to delta function self-correlations of the environment.
\\
\\
3. We find that on sub-horizon scales the density matrix violates positivity, a symptom of the fact that the master equation is not in Lindblad -- the manifestly positivity preserving -- form. This shows that on sub-horizon scales the assumptions that go into deriving the master equation cannot be completely consistent. However, positivity is recovered on super-horizon scales when the state becomes very squeezed. We also see the same behaviour at the level of the Wigner function, which satisfies a Fokker-Planck equation exhibiting unphysical ``negative-diffusion" on sub-horizon scales which becomes unimportant as modes exit the horizon. In this sense we provide a complete picture of how decoherence and positivity emerge as modes exit the horizon leading to the interpretation of the density matrix as a classical Probability Density Function (PDF) after it becomes both diagonal and positive.
\\
\\
4. It is a well-known fact that in realistic systems, entanglement between spacelike separated regions is extremely sensitive to noise from the local environment \cite{BP}, it is therefore important to ask how vulnerable entanglement produced during inflation is to decoherence. The final part of this paper uses the quantum master equation and its solutions to study the effects of environmental noise on entanglement produced during inflation. The principle measure we use is quantum discord \cite{OZ, HV}. Our analysis reveals that the environment does indeed weaken the strength of discord with the corrections being of order $\lambda/H$, a ratio characterising the relative strength of entangled particle pair production and the interaction rate with the environment.
\\
\\
\indent The structure of the remainder of the paper is as follows. Beginning in sec.~\ref{philosophy}, we discuss open quantum systems in general terms using the paradigm of quantum Brownian motion to describe various concepts, which is particularly fitting as the mode-by-mode master equation of cosmological perturbations assumes the same schematic form. Then, in sec.~\ref{masterEq}, we introduce the simplified model used to study an open system of primordial perturbations and review the derivation of the corresponding master equation outlined in \cite{Boyanovsky:2015jen}.
\\
\\
Once this groundwork is laid, in sec.~\ref{interpretation} we provide a detailed analysis of the master equation, drawing comparisons with quantum Brownian motion \cite{BP,CL,H}, where we identify the usual fluctuation-dissipation terms, in addition to the unitary coherent evolution associated to the isolated squeezed state. We find that, although the master equation is not in Lindblad form, positivity of the density matrix is established on super-horizon scales, providing a natural emergence of classicality at late times. We also write the master equation in terms of the Wigner function which we then use to construct a Langevin-type equation. This provides an equivalent description of the model -- in the sense of expectation values -- in terms of a stochastic theory of classical trajectories. In this sense we provide a first-principles derivation in the context of the model \eqref{interaction}, of a classical stochastic equation which reproduces the dynamics of the quantum master equation on super-horizon scales. It should be noted that this is different from the stochastic inflation paradigm \cite{Star1, SY}, where the noise is described by sub-Hubble modes in contrast to the present setup, where the noise arises from another field, conformally coupled so that it can mimic the sub-Hubble environment in a computational tractable model \cite{Boyanovsky:2015jen}.

Finally, in sec.~\ref{discordSec} we shall address the question of whether quantum correlations (generated by entangled pair creation) are robust to environmental decoherence. Recently, a novel approach to entanglement in inflationary cosmology has been explored by Lim \cite{Lim} and Vennin and Martin \cite{VM}. The idea is to harness the power of correlations in the system of perturbations using \textit{quantum discord} \cite{OZ,HV} which relies on the difference between classical and quantum correlations to quantify the extent to which a particular system exhibits quantum features.
\\
\\
It is natural to expect entanglement between different regions of a system to be subdued due to entanglement with a local environment. Indeed it has already been noted in other contexts that an external environment weakens discord \cite{BA}. The detailed analysis of decoherence provided by our quantum master equation and its solutions allows us to examine precisely this question. We find that environmental decoherence has a tendency to reduce the size of the discord and offer an interpretation of this in terms of the competing effects of particle pair creation in de Sitter (which tends to increase the strength correlations) and random production of particles due to the environmental scatterings, which weakens the strength of correlations. In section \ref{discussion} we offer our conclusions and some suggestions for future work.

\section{The quantum master equation}\label{philosophy}

Since our central goal to is understand how a quantum state can end up classical, the nature of the quantum-to-classical transition deserves some comment. We take a pragmatic view here: classical mechanics is manifestly an emergent phenomenon of quantum systems in certain circumstances and the two key issues are: (i) what are those circumstances? And, (ii) what are the phenomenological rules that describe how the classical state emerges from an underlying quantum state?

We view the emergence of classical mechanics in the same way one views any effective theory of some underlying microscopic system. The resulting description will always be phenomenological and only appropriate to a coarse-grained perspective. By ``coarse-grained" we mean that the effective theory is couched in terms of observables that have some realistic finite resolution scales, e.g.~the dynamics of fluctuations on superhorizon scales in the cosmological setting. For observers probing these scales, an evolving quantum state can be equivalent to an ensemble of classical stochastic trajectories in the sense that quantum expectation values are captured by stochastic averaging. The final step is to postulate that an individual classical stochastic trajectory is {\it real\/} for the coarse-grained observer.

\subsection{Analogy with quantum Brownian motion}

A useful toy model which has many parallels with the present investigation is the much-discussed subject of quantum Brownian motion \cite{H,BP,CL,JZ}. The system consists of a particle, which, in the simplest setup, moves in one dimension, interacting with an environment. The environment can be modelled in many different ways leading to the same universal behaviour for the particle. The first level of coarse graining is to take a perspective -- defined by a set of observables -- which act on the particle alone, e.g.~its position and momentum. In these circumstances, one can trace out the environment and work with the reduced density matrix of the particle, $\rho=\text{Tr}_\mathcal{E}|\Psi\rangle\langle\Psi|$. Here, $\ket{\Psi}$ is the state vector of the joint environment-particle system whose evolution is unitary. In general, $\rho$ does not satisfy an autonomous dynamical equation; however, there are circumstances for which one can derive an \textit{approximate} autonomous equation for $\rho$ known as the ``master equation". In going from the unitary evolution of the total state, to non-unitary evolution encoded in the quantum master equation, there are essentially three stages of approximation which must be made and justified:
\begin{enumerate}
\item The first is the Born approximation, in which one replaces the total (pure) state of the system $|\Psi\rangle$ by a product of density matrices for the particle and environment $\rho_S\otimes \rho_\mathcal{E}$. The idea is that the environment is so large compared with the system (the particle) that in every instant the system is interacting with a fresh bit of the environment and the overall environment is barely affected by the interaction with the much smaller system. Any entanglement that previously builds up between the particle and the environment is encoded in the von Neumann entropy of the reduced density matrix of the particle which increases over time.
\item The second approximation is the Markov approximation that allows one to replace terms that depend on the history of the system by the instantaneous state $\rho$. This approximation is justified when the dynamics of $\mathcal{E}$ is suitably fast compared with the slow dynamics of the macroscopic particle.
\item Finally, the resulting equation for $\rho$ has terms that oscillate rapidly on the fast environmental time scale. Temporal coarse graining amounts to removing these terms by hand in a form of rotating wave approximation \cite{BP}.
\end{enumerate}
If all these steps are followed, a master equation for $\rho$ can emerge with solutions that are physically consistent, i.e.~which preserve the positivity of the spectrum of $\rho$. The latter is needed for a consistent density matrix which must have a spectrum $\{p_n\}$ for which $p_n\geq0$ and $\sum_np_n=1$ (the normalization condition). The necessary and sufficient condition for this is that the master equation can be written in Lindblad form, that is as
\EQ{
\dot\rho=\frac1{i\hbar}[H,\rho]+\sum_i\alpha_i\Big(2A_i\rho A_i^\dagger-A^\dagger_iA_i\rho-\rho A^\dagger_iA_i\Big)\ ,
\label{lin}
}
with $\alpha_i\in\mathbb R>0$. It is the second term here that encodes the interaction with the environment.
For quantum Brownian motion, the form of the master equation in the position basis $\rho(x,x')=\langle x|\rho|x'\rangle$, is the Caldeira-Leggett master equation \cite{CL}.
\EQ{
\dot\rho&=\overbracket{\frac1{i\hbar}[H_\text{eff},\rho]}^{\text{unitary evol.}}-\overbracket{\gamma(x-x')\Big(\frac\partial{\partial x}-\frac\partial{\partial x'}\Big)\rho}^{\text{dissipation}}\\ &\qquad\qquad-\underbracket{\frac{2m\gamma k_BT}{\hbar^2}(x-x')^2\rho}_{\text{decoherence}}\ .
\label{mse}
}
Here, $H_\text{eff}$ is the Hamiltonian of the particle with some additional additive terms that arise from the coupling of the particle and $\mathcal{E}$. In fact, strictly speaking, this master equation is not in Lindblad form. To render it so, one must take
\EQ{
A=x+i\beta p\ ,\qquad \alpha=\frac{2m\gamma k_BT}{\hbar^2}\ ,\qquad\beta=\frac\gamma\alpha\ ,
}
in the Lindblad form \eqref{lin}, and this gives rise to extra terms quadratic in momentum.
However, in realistic situations where $T$ is not too small, the departure from positivity is small at the level of solutions, so the lack of terms needed for positivity in \eqref{mse} can be neglected. The characteristic features of the master equation are the dissipation (or relaxation) and fluctuation (decoherence) terms.

Now that we have an explicit example of an open quantum system, we can ask the all-important question: what makes a state of this system ``classical"? To start with the word ``classical" is attached to quantum states in different ways. One answer is that a classical state is one which is highly localised in position and momentum, like a coherent state (i.e.~a state which provides the minimal level of uncertainty allowed by the Heisenberg relation). If the spread $\delta x$ is much smaller than the scale over which the potential energy $V(x)$ varies then Ehrenfest's theorem guarantees that the quantum state follows the classical trajectory. Unfortunately, this reasoning, although often invoked in textbooks, is completely unrealistic. Even in free systems, wave functions spread over time. This spreading is slow for a macroscopic particle but in any chaotic system this spreading becomes remarkably fast, order $t_c\log(I/\hbar)$, where $t_c$ is the characteristic (classical) chaos time scale and $I$ is the characteristic (macroscopic) action scale of the system. In summary, we cannot rely on Ehrenfest's theorem to describe the quantum-to-classical transition (see the discussion in \cite{Zurek:1998mj}).

The second use of the word ``classical" is in ``semi-classical", or WKB states. These kinds of state are pure and are smeared along classical trajectories of the system in phase space and so are not the kind of state that can describe the quantum-to-classical transition \cite{Berry}.

Finally, we have the notion of a classical state that is meaningful for the quantum-to-classical transition. These states are very spread out in both position and momentum. This seems paradoxical: how can such states be classical? It will emerge that the answer is that these states give quantum expectation values that can be captured by a classical stochastic process.

A state which is very spread out in position and momentum has
coherence lengths $\ell_x$ and $\ell_p$ (the scales over which the off-diagonal elements of the density matrix fall off in position and momentum space, respectively) which are very small in the sense that
\EQ{
\ell_x\ell_p\ll\hbar\ .
\label{pip}
}
Such states must necessarily be mixed states rather than pure and can be given a useful phase space representation by using, for example, the Wigner function. We choose the latter for its ubiquity even though there is nothing that really singles it out from other alternatives (e.g.~the Husimi function):
\EQ{
W(x,p)=\frac1\pi\int_{-\infty}^\infty dx'\,\rho\Big(x-\frac{x'}2,x+\frac{x'}2\Big)e^{2ipx'/\hbar}\ .
\label{wfn}
}
In general, there is nothing to guarantee the positivity of $W(x,p)$ -- unlike the Husimi function which is manifestly positive -- but for a ``classical state" spread out in phase space over an area $\gg\hbar$, $W(x,p)$ becomes positive.

The master equation \eqref{mse} can be written in terms of the Wigner function as
\EQ{
\dot W&=\overbracket{\{H_\text{eff},W\}}^{\text{Poisson bra.}}+
\overbracket{2\gamma\partial_p(pW)}^{\text{dissipation}}+\overbracket{2m\gamma k_BT\partial_p^2W}^{\text{decoherence}}\\
&+\sum_{n\geq 1}\underbracket{\frac{(-1)^n\hbar^{2n}}{2^{2n}(2n+1)!}\partial_x^{2n+1}V\partial_p^{2n+1}W}_{\text{quantum corrections}}
\ .
\label{wme}
}
Note that the quantum corrections above arise from non-linear terms in the dynamics, that is, on the derivatives of the potential $V(x)$ of order three and greater.

We can now consider the effect of the various terms in the master equation. Firstly, with only the Poisson bracket term on the right-hand side, Liouville's theorem would ensure that the area of the support of $W$ is preserved. Consider an initial state with minimal support allowed by the uncertainty principle $A\sim\hbar$ -- a coherent state. The non-linear interactions in the Poisson bracket term have the tendency to make $W$ spread out in phase space but at the same time vary on small scales so that the area remains fixed $A\sim\hbar$. Generally the area is squeezed in some directions and expands in other directions in such a way that the perimeter grows rapidly. Now consider the effect of the quantum corrections in \eqref{wme}. These generally destroy the probability density interpretation of $W$ and encode non trivial quantum interference. Specifically, these terms will lead to a violation of the positivity of the Wigner function as it ``interferes with itself".

However, the decoherence terms coming from the interactions of the particle with the environment can rescue a probabilistic interpretation. In particular the decoherence term, causes the Wigner function to spread out but also become smooth on small scales. The positivity violations are alleviated and the area of support is driven to $A\gg\hbar$. This spreading and smoothing goes hand in hand with coherence lengths that are driven to the regime \eqref{pip}. This lack of purity can also be quantified by the entanglement entropy. Note that the smoothing out caused by decoherence also implies that the different phase space representations of a state, Wigner, Husimi, etc., essentially become indistinguishable.

The fact that the Wigner function becomes smoothed out means that the derivatives $\partial_p^nW$ are small and so the quantum corrections in \eqref{wme} are suppressed \cite{Zurek:1998mj}. Effectively one can describe the dynamics of the Wigner function by the much simpler equation
\EQ{
\dot W\approx\{H_\text{eff},W\}+
2\gamma\partial_p(pW)+2m\gamma k_BT\partial_p^2W\ .
\label{wme2}
}
Now comes the key point for the quantum-to-classical transition. This simplified master equation has precisely the form of a Fokker-Planck equation where the decoherence term corresponds to diffusion and the dissipation term to friction. In other words, the Wigner function can now be given the interpretation of a classical probability density function for a set of classical stochastic trajectories described by the Langevin equations
\EQ{
\dot x=\{x,H_\text{eff}\}\ ,\qquad
\dot p=\{p,H_\text{eff}\}-2\gamma p+\sigma\ ,
}
where $\sigma(t)$ is a Gaussian white noise term with characteristics
\EQ{
{\mathbb E}(\sigma(t))=0\ ,\qquad {\mathbb E}(\sigma(t)\sigma(t'))=4m\gamma k_B T\delta(t-t')\ .
}

The final step in the quantum-to-classical transition is to interpret the Langevin equation as describing classical trajectories that have {\it emerged out\/} of the underlying quantum dynamics at the level of the coarse-grained perspective (after tracing over the environment). Note that the equivalence to a classical stochastic process means that quantum expectation values can be captured by averages over classical stochastic trajectories:
\EQ{
\text{Tr}\big(\rho(t){\cal O}(x,p)\big) \approx \sum_{\{x(t'),p(t')\}}{\cal O}(x(t),p(t))\ ,
}
where the average on the right-hand side is over the stochastic trajectories up to time $t$ with random initial conditions that correspond to the initial Wigner function.

Let us summarize the phenomenological approach to the quantum-to-classical transition:
\begin{center}
\begin{tikzpicture}[scale=1]
\node at (0,1) (a1){$H|\Psi\rangle=i\hbar\partial_{t}|\Psi\rangle$};
\node at (0,0) (b1) {coarse grain over $E$};
\node at (0,-1) (a2) {master equation for $\rho$};
\node at (0,-2) (b2) {decoherence};
\node at (0,-3) (a3) {Effective Fokker-Planck equation};
\node at (0,-4) (a4) {Langevin equation for emergent classical trajectories};
\draw[very thick] (a1) -- (b1);
\draw[very thick,->] (b1) -- (a2);
\draw[very thick] (a2) -- (b2);
\draw[very thick,->] (b2) -- (a3);
\draw[very thick,->] (a3) -- (a4);
\draw[very thick] (-1.7,0.24) -- (1.7,0.24) -- (1.7,-0.24) -- (-1.7,-0.24) -- (-1.7,0.24);
\draw[very thick] (-1.3,-1.76) -- (1.3,-1.76) -- (1.3,-2.24) -- (-1.3,-2.24) -- (-1.3,-1.76);
\end{tikzpicture}
\end{center}
We shall now see how this same chain of steps emerges during inflation, with environmental interactions leading to diffusion/decoherence terms in the master equation that are then interpreted as stochastic noise acting on the emergent classical trajectories. The reduced dynamics is given by tracing out this environment with an additional level of coarse graining provided by restricting to super-Hubble modes -- the appropriate scales for analysing the CMB.

It is worth pointing out at this stage, that what we have described above is a way to think about the quantum-to-classical transition and how classical trajectories can be thought to have emerged from the quantum system in an interacting system. However, the simplest cosmological model is free, i.e.~$V(x)$ is quadratic and there is no environment. In that case, the master equation for the Wigner function obviously takes the form of a Fokker-Planck equation immediately and therefore is clearly equivalent to a classical stochastic process where the trajectories in this case satisfy the classical equations of motion (there is no noise in this case because there is no diffusion term) but with initial conditions that are defined randomly with respect to a PDF that is the initial Wigner function (this is shown quite explicitly in \cite{KP,Kiefer:2008ku,Martin:2012pea}). Note that this interpretation can be made even though the quantum state remains pure. It is really a moot point as to whether this as a valid description of the quantum-to-classical transition because in any realistic system, including the cosmological one, we expect that there will be interactions that cannot be ignored.

\section{Master equation for the reduced density matrix}\label{masterEq}

In order to study an open system of the scalar cosmological perturbations, we introduce a simplified model considered by Boyanovsky \cite{Boyanovsky:2015jen} (whose notation we follow) by coupling the scalar perturbations $\phi$ to an external scalar field $\varphi$ via the action
\begin{align}
S &= \int d^3 x \, dt \, a^3\, \Big\{ \frac{1}{2} \dot{\phi}^2 - \frac{(\nabla \phi)^2}{2 a^2} - \frac{1}{2}\left(M_\phi^2 + \xi_\phi \mathcal{R} \right) \phi^2 \nonumber \\
&+\frac{1}{2} \dot{\varphi}^2 - \frac{(\nabla \varphi)^2}{2 a^2} - \frac{1}{2}\left(M_\varphi^2 + \xi_\varphi\mathcal{R} \right) \varphi^2 - \lambda \, \phi \, : \varphi^2 :\Big \},
\end{align}
where
\begin{equation}
\mathcal{R} = 6 \left( \frac{\ddot{a}}{a} + \frac{\dot{a}^2}{a^2} \right),
\end{equation}
is the Ricci scalar and $\xi=0,1/6$ correspond to the cases of minimal and conformal coupling, respectively. The interaction is normal-ordered so that
\begin{equation}
: \varphi^2: = \varphi^2 - \braket{\varphi^2},
\end{equation}
where $\langle\cdots\rangle$ denotes a QFT correlator.
We shall work in conformally rescaled fields $\phi = \chi/a$ and $\varphi = \psi/a$ and introduce conformal time $\eta = -1/aH$, in which the action becomes
\EQ{
S &= \int d^3 x \, d \eta\, \Big\{ \frac{1}{2}\big[ \chi'^2 - (\nabla \chi)^2 - \mathcal{M}^2_\chi(\eta) \chi^2 \big] \\
& +\frac{1}{2}\big[ \psi'^2 - (\nabla \psi)^2 - \mathcal{M}^2_{\psi}(\eta) \psi^2 \big] + \frac{\lambda}{H \eta} \, \chi : \psi^2:
\Big\},
\label{action}
}
where
\begin{equation}
\mathcal{M}_{\chi, \psi}(\eta) = \Big[\frac{\mathcal{M}_{\chi, \psi} ^2 }{H^2} + 12 \Big( \xi_{\chi, \psi} - \frac{1}{6} \Big) \Big] \frac{1}{\eta^2},
\end{equation}
provides an effective, time-dependent mass.

In order to provide a simplified model of the cosmological perturbations, $\chi$ plays the r\^ole of the Mukhanov-Sasaki variable and so is a minimally coupled ($\xi_\chi=0$) massless field. In fact, the bare mass will be non-vanishing in order to absorb a UV divergence.
On the other hand, for the environment field, choices can be made. We follow \cite{Boyanovsky:2015jen} and take $\psi$ to be a conformally coupled field ($\xi_{\psi} = 1/6$). The intuition for this choice is that the field $\psi$ can either be viewed as a field in its own right or as a proxy for the high frequency components of $\chi$ whose wavelength remain inside the horizon during inflation. This mimics a non-linear self-interaction that couples modes of different momentum. In this sense, the simplified model attempts to model an IR-UV split of the degrees of freedom of a single (interacting) field.

It is convenient -- at least to start with -- to work in the interaction picture where the observables evolve in the free Heisenberg picture and the density matrix evolves according to the von Neumann equation with the interaction Hamiltonian
\begin{equation}\label{drhoI}
\frac{d \rho_I (\eta)}{d \eta} = -i \left[H_I(\eta), \rho_I(\eta) \right].
\end{equation}
where
\begin{equation}\label{interaction2}
H_I(\eta) = - \frac{\lambda}{H \eta} \int d^3 x\, \chi(\BX, \eta) : \psi^2(\BX, \eta):,
\end{equation}
and where $\chi(\BX, \eta)$ and $\psi(\BX, \eta)$ are fields in the free Heisenberg representation. It is worth pointing out at this stage that since the interaction (\ref{interaction}) is time dependent, na\"\i ve perturbation theory is only valid on timescales satisfying $\lambda/H \lesssim |\eta|$.

The von Neumann equation \eqref{drhoI} has solutions
\begin{equation}
\rho_I(\eta) = \rho_I(\eta_0) - i \int^\eta_{\eta_0} d \eta' \left[H_I(\eta'),\rho_I(\eta') \right].
\end{equation}
Feeding this back into (\ref{drhoI}) gives the leading terms in the perturbative expansion describing the evolution of $\rho_I$:
\EQ{
\frac{d \rho_I (\eta)}{d \eta} & = -i \left[H_I(\eta), \rho_I(\eta_0) \right] \\
&-\int^\eta_{\eta_0} d \eta' \left[ H_I(\eta), \left[H_I(\eta'),\rho_I(\eta') \right] \right].\label{drhoevolution}
}

In order to derive an autonomous equation for the evolution of the density matrix of the field $\chi$ -- the master equation -- some familiar approximations are necessary. Firstly, the Born approximation assumes that the total density matrix can effectively be factorised as
\begin{equation}
\rho_I(\eta) = \rho_{I \chi} (\eta) \otimes \rho_{ I \psi} (\eta_0).
\label{factorisation}
\end{equation}
As described in sec.~\ref{philosophy}, this is justified when the environment is large enough such that in each time interval, the system effectively interacts (weakly) with a fresh part of the environment. Essentially, the environment is unaffected by the interaction, while the system's entanglement with the environment is encoded in the reduced density matrix $\rho_{I\chi}$. The reason why factorisation may be justified in the present context is due to the fact that a single mode of the field $\chi$ couples to a pair of modes of the environmental scalar field and in a de Sitter background there are no particle thresholds.
Brownian motion, described in sec.~\ref{philosophy}, provides a familiar example which exploits these same approximations. In the present context, this approximation is described in greater detail in \cite{BP,Boy1, Boyanovsky:2015jen}.

Combining the approximations (\ref{drhoevolution}) and (\ref{factorisation}) leads to the following equation \cite{Boyanovsky:2015jen} for the reduced interaction-picture density matrix $\rho_{I\chi}(\eta)$ (in the following we will now denote the reduced density matrix of $\chi$: $\rho_I\equiv\rho_{I\chi}$):
\begin{widetext}
\begin{align}\label{masterposition}
\frac{d\rho_I(\eta)}{d\eta}& = \frac{- \lambda^2}{H^2 \eta} \int^\eta_{\eta_0} \frac{d \eta'}{\eta'} \int d^3 x \int d^3 y \Big\{
\chi(\BX,\eta)\, \chi(\BY ,\eta') \, \rho_I(\eta') \, G^>(\BX - \BY , \eta, \eta') + \rho_I(\eta')\, \chi(\BY ,\eta')\, \chi(\BX,\eta)\, \, G^<(\BX - \BY , \eta, \eta') \nonumber \\
& -\, \chi(\BX,\eta')\, \rho_I(\eta') \, \chi(\BY ,\eta)\, \, G^<(\BX - \BY , \eta, \eta') -\, \chi(\BY ,\eta) \rho_I(\eta') \chi(\BX,\eta')\, \, G^>(\BX - \BY , \eta, \eta')
\Big\},
\end{align}
where
\EQ{
G^>(\BX - \BY , \eta, \eta') &= \text{Tr} \left[ :\psi^2 (\BX, \eta) : :\psi^2 (\BY , \eta) : \rho_{I \psi} (\eta_0) \right], \\
G^<(\BX - \BY , \eta, \eta') &= \text{Tr} \left[ :\psi^2 (\BY , \eta) : :\psi^2 (\BX, \eta) : \rho_{I \psi} (\eta_0) \right],
}
are environmental correlators.

Our next step is to make the Markov approximation by taking $\rho_I(\eta') \rightarrow \rho_I(\eta)$ in the integral expression in eq.~(\ref{masterposition}). This is consistent with the weak coupling limit, as can be seen with an integration by parts of (\ref{masterposition}), noting that $d\rho_I/d\eta \propto \lambda^2/H^2 \ll 1$ and neglecting the resulting term of $\mathcal{O}(\lambda^4)$. Going to momentum space, we write
\begin{align}
G^\lessgtr(\BX - \BY , \eta, \eta') &= \frac{1}{V} \sum_{\Bk } \mathcal{K}^\lessgtr(k, \eta, \eta')e^{- i \Bk \cdot(\BX- \BY )},\label{Ggtr}
\end{align}
where V is the quantisation volume, an IR regulator.

In order to proceed, we must compute the objects $K^{\lessgtr}(p,\eta,\eta')$. As shown in \cite{Boyanovsky:2015jen}, after specialising to a Bunch-Davies (BD) vacuum \cite{BD} for the initial conditions of the field $\psi$,
$\rho_{I\psi}(\eta_0) = \tensor[]{\ket{0}}{_\psi} \! \! \tensor[_\psi]{\bra{0}}{} $, we find
\begin{align}
\mathcal{K}^>(q,\eta,\eta') \equiv K(q,\eta,\eta')= 2 \int \frac{d^3 k }{(2\pi)^3} v(k,\eta) v^*(k,\eta') v(p,\eta) v^*(p,\eta')
\label{Kp}
\end{align}
and ${\cal K}^<(q,\eta,\eta')=K^*(q,\eta,\eta')$, where $ q = |\Bk + \BP |$ and $v(k,\eta)$, \textit{etc}., are the mode functions associated to the environment field $\psi$. They satisfy
\begin{equation}
v'' + \Big[k^2 - \frac{1}{\eta^2} \Big(\nu_\psi^2 - \frac{1}{4} \Big)\Big] v = 0,
\end{equation}
with $\nu^2_\psi = 9/4 - (M_\psi^2/H^2 + 12 \xi_\psi)$. The steps above lead to a master equation
\EQ{\label{rhoX}
\frac{d\rho_I(\eta)}{d\eta} = - \frac{\lambda}{H^2 \eta} \int^\eta_{\eta_0} \frac{d \eta'}{\eta'} \sum_{\Bk }
&\Big\{
\chi_{\Bk }(\eta)\, \chi_{-\Bk }(\eta') \rho_I(\eta) K(k, \eta, \eta') +\rho_I(\eta) \, \chi_{- \Bk }(\eta')\, \chi_{\Bk }(\eta) K^*(k, \eta, \eta') \\
&- \chi_{\Bk }(\eta) \rho_I(\eta) \, \chi_{- \Bk }(\eta')\, K^*(k, \eta, \eta') - \chi_{- \Bk }(\eta) \rho_I(\eta) \, \chi_{\Bk }(\eta)\, K^*(k, \eta, \eta')
\Big\},
}
where $\chi_\Bk (\eta)$ are the Fourier modes of the field satisfying $\chi^\dagger_{\Bk }= \chi_{-\Bk }$.
\end{widetext}

Since $\psi$ is massless and conformally coupled, with BD vacuum initial conditions, it has simple mode functions
\begin{equation}
v(k,\eta) = \frac{e^{-i k \eta}}{\sqrt{2k}}.
\end{equation}
In this case,
\begin{equation}\label{K}
K(p, \eta,\eta') = - \frac{i}{8 \pi^2} e^{- i p (\eta - \eta')} \mathcal{P}\left[ \frac{1}{\eta - \eta'} \right] + \frac{1}{8 \pi}\delta(\eta - \eta'),
\end{equation}
where $\mathcal{P}$ stands for the principal part, which can be written as
\begin{align}\label{Principal}
\mathcal{P} \left[\frac{1}{\eta - \eta'} \right] = -\frac{1}{2}\frac{d}{d \eta}\log \left[ \frac{(\eta - \eta')^2 + \varepsilon^2}{(- \bar{\eta})^2 } \right].
\end{align}
Here $\varepsilon\to0$ is a UV regulator and $\bar{\eta}$ is an arbitrary scale to make the argument of the logarithm dimensionless, and, as we shall see shortly, it also acts as a renormalisation scale. Inserting eqs.~(\ref{K}) and (\ref{Principal}) into eq.~(\ref{rhoX}) leads to terms of the form
\begin{align}
&\int \frac{d \eta'}{\eta'} \chi_{-\Bk }(\eta') K[q; \eta, \eta'] \nonumber \\
& \equiv - \frac{1}{2}\int^\eta_{\eta_0} \! \! d \eta' \, \, \chi_{-\Bk} \frac{e^{- i k (\eta - \eta')}}{\eta'}\frac{d }{d \eta'} \log \left[\frac{(\eta - \eta')^2 + \varepsilon^2}{(- \bar{\eta})^2} \right].
\end{align}
Integrating by parts yields
\EQ{\label{non-local}
& \left. - \frac{1}{2} \, \, \chi_{-\Bk }(\eta') \frac{e^{- i k (\eta - \eta')}}{\eta'} \log \left[\frac{(\eta - \eta')^2 + \varepsilon^2}{(- \bar{\eta})^2} \right] \right|_{\eta'=\eta_0}^{\eta'=\eta} \\
& + \int^\eta_{\eta_0} \! \! d \eta' \, \, \frac{d }{d \eta'}\left[ e^{- i (\eta - \eta')} \frac{\chi_{-\Bk}(\eta')}{\eta'}\right] \log \left[\frac{\eta - \eta'}{- \bar{\eta}} \right].
}
As explained in \cite{Boyanovsky:2015jen} by identifying the RG scale with $\eta_0$ and by taking $- k \eta_0 \gg1$ we can neglect the lower end of the surface term. The upper end gives a UV divergence which is removed by an additive mass renormalisation of $\chi$.

These steps lead to a master equation consisting of local and non-local terms
\begin{equation}\label{Dequation}
\frac{d\rho_I(\eta)}{d\eta} = \mathcal{D}_L [\rho_I] + \mathcal{D}_{NL}[\rho_I],
\end{equation}
where
\EQ{
&\mathcal{D}_{L}[\rho_I] = - \frac{i\lambda^2}{8\pi^2 H^2 \eta}\log\Big[\frac{\varepsilon}{-\eta_0}\Big] \sum_{\BP }
\left[
\chi_\BQ (\eta) \chi_{-\BP }(\eta), \rho_I(\eta) \right] \\
& - \frac{\lambda^2}{16 \pi H^2 \eta^2}\sum_{\BP }
\Big\{
\chi_{ \BP }(\eta) \chi_{ -\BQ }(\eta) \rho_I(\eta)\\
&+ \rho_I(\eta) \chi_\BP (\eta) \chi_{-\BP }(\eta) - 2 \chi_{-\BP }(\eta) \rho_I(\eta) \chi_{\BP }(\eta)
\Big \},
}
and,
\EQ{
\mathcal{D}_{NL}[\rho_I] & = \frac{i \lambda^2}{8 \pi H^2 \eta^2}\sum_{\BP }
\Big\{
\chi_{ \BP }(\eta) X_{ -\BQ }(\eta) \rho_I(\eta) \\
& -\rho_I(\eta) \overline{X}_{-\BP }(\eta) \chi_{\BP }(\eta) + \chi_{\BP }(\eta) \rho_I(\eta) \overline{X}_{-\BP }(\eta) \\
&- X_{-\BP }(\eta) \rho_I(\eta) \chi_{\BP }(\eta) \Big \},
\label{NLterm}}
with
\EQ{\label{non-loca2}
X_{- \Bk }(\eta) = \int^\eta_{\eta_0} \! \! d \eta' \, \, \frac{d }{d \eta'}\Big[ e^{- i k (\eta - \eta')} \frac{\chi_{-\Bk}(\eta')}{\eta'}\Big] \log \Big[\frac{\eta - \eta'}{- \eta_0} \Big]
}
and
\EQ{\label{non-localBAR}
\overline{X}_{-\Bk }(\eta) = \int^\eta_{\eta_0} \! \! d \eta' \, \, \frac{d }{d \eta'}\Big[ e^{ i k (\eta - \eta')} \frac{\chi_{-\Bk }(\eta')}{\eta'}\Big] \log \Big[\frac{\eta - \eta'}{- \eta_0} \Big].
}
The $\varepsilon$ divergence in the local term is removed by mass renormalisation of $\chi$. Notice that the non-local contribution arises as a result of integrations over long-time environment correlations, and is therefore associated with memory effects. In contrast, the local terms are Markovian, and in the stochastic interpretation that we will develop, corresponds to white noise. This distinction will prove important in our subsequent analysis. This completes our derivation of the master equation and review of the main steps in \cite{Boy1,Boyanovsky:2015jen}.

\section{Interpreting the master equation}\label{interpretation}

We begin by rewriting the master equation in the Schr\"odinger picture, which allows us to draw close parallels with quantum Brownian motion. To do this, we rewrite the non-local contribution in terms of Heisenberg operators at time $\eta$ by writing the modes $\chi_{-\Bk }(\eta')$ in the integral kernel as
\begin{equation}
\chi (\eta') = f(\eta', \eta) \chi (\eta) +k^{-1} g(\eta',\eta) \Pi(\eta),
\label{pop}
\end{equation}
where $f$ and $g$ are functions to be determined. For now we have suppressed the momentum labels $\Bk$. The factor $k^{-1}$ is there on dimensional grounds and will drop out of our calculations at a later stage. Notice in particular that the Heisenberg equations of motion for $\chi(\eta')$ imply that $f$ and $g$ must satisfy the mode equation with respect to $\eta'$; namely
\begin{equation}
f'' + \left(k^2 - \frac{a''}{a} \right)f =0,
\end{equation}
and similarly for $g$. Furthermore, the relation
\begin{equation}
\Pi = \chi' - \frac{a'}{a},
\end{equation}
means that
\begin{equation}
\Pi(\eta') = \Big(f' - \frac{a'}{a}f \Big)\chi (\eta) +k^{-1} \Big(g' - \frac{a'}{a} g \Big) \Pi(\eta).
\end{equation}
Noting that $a'/a=-1/\eta$, we see that $f$ and $g$ must satisfy the boundary conditions
\EQ{
&f(\eta'=\eta)=1, \quad \partial_{\eta'}f(\eta'=\eta)=-1/\eta,\\
&g(\eta'=\eta)=0, \qquad \partial_{\eta'}g(\eta'=\eta)=k.
}
Solving explicitly for the mode functions, we find
\EQ{
g\left(\eta',\eta\right) &= \Big( \frac{1}{k \eta'}-\frac{1}{k \eta}\Big) \cos\big[ k \left( \eta-\eta'\right)\big] \\
& - \Big(1+\frac{1}{k^2 \eta \eta'}\Big) \sin \big[ k \left( \eta-\eta'\right)\big]\label{g},
}
and
\begin{align}
f(\eta',\eta) & =\frac{\sin \left[ k \left( \eta-\eta'\right)\right]}{k \eta'}+\cos\left[ k \left( \eta-\eta'\right)\right] \label{f}.
\end{align}
With these definitions we find that the non-local objects $X_{-\Bk }(\eta)$ in (\ref{NLterm})-(\ref{non-localBAR}) can be written as a linear combination of Heisenberg operators $\chi(\eta)$ and $\Pi(\eta)$, multiplied by integrals over the functions $f(\eta',\eta)$ and $g(\eta',\eta)$, \textit{i.e.},
\begin{equation}
X_{-\Bk}(\eta ) = k \, \chi_{-\Bk}(\eta) F(\eta) + \Pi_{-\Bk}(\eta) G (\eta)
\end{equation}
where
\begin{align}
F(\eta) & = \frac{1}{k} \int_{\eta_0}^\eta d\eta' \, \, \frac{d}{d \eta'}\Big[ f(\eta',\eta) \frac{ e^{-i k (\eta-\eta')}}{\eta'}\Big] \log \Big[\frac{\eta - \eta'}{- \eta_0} \Big] , \label{F} \\[5pt]
G(\eta) & = \frac{1}{k} \int_{\eta_0}^\eta d\eta' \, \, \frac{d}{d \eta'}\Big[ g(\eta',\eta) \frac{ e^{-i k (\eta-\eta')}}{\eta'}\Big] \log \Big[\frac{\eta - \eta'}{- \eta_0} \Big],\label{G}
\end{align}
and correspond to long-time correlations in the environment. We can now perform the switch to the Schr{\"o}dinger picture. Using the result
\begin{equation}
U(\eta,\eta_0) \chi_\Bk (\eta) U^{-1}(\eta,\eta_0) = \chi_\Bk
\end{equation}
and the relation between the interaction density matrix and the (time-dependent) Schr\"odinger picture density operator
\begin{equation}
\rho_I(\eta) = U_0^{-1}(\eta,\eta_0) \rho(\eta) U_0(\eta,\eta'),
\end{equation}
as well as
\begin{equation}
\frac{d U_0(\eta,\eta_0) }{d \eta} = - i \left[H_0(\eta),\rho \right],
\end{equation}
we find that after appropriate insertions of $U U^{-1}$ in the standard way, one obtains a Schr{\"o}dinger representation of the master equation
\begin{widetext}
\EQ{\label{drho2}
\frac{d \rho(\eta)}{d \eta}& = -i [H_0,\rho(\eta)] - \frac{\lambda^2}{16 \pi H^2 \eta^2} \sum_{\Bk }
\Big\{\chi_{\Bk } \chi_{-\Bk } \rho(\eta) +\rho (\eta) \chi_{\Bk }\chi_{-\Bk } - 2\chi_{-\Bk } \rho (\eta) \chi_{\Bk } \Big\}
\\
&+ \frac{i k\lambda^2 }{8 \pi^2 H^2 \eta} \sum_{\Bk }
\Big\{
F(\eta) \left[ \chi_{\Bk } \chi_{-\Bk } \rho(\eta)- \chi_{-\Bk }\rho(\eta) \chi_{\Bk }\right]
- F^*(\eta) \left[ \rho(\eta) \chi_{- \Bk } \chi_{\Bk } - \chi_{\Bk }\, \rho(\eta) \chi_{-\Bk }\right]\Big \} \\
&+ \frac{i\lambda^2 }{8 \pi^2 H^2 \eta} \,\sum_{\Bk }
\Big\{
G(\eta)\left[ \chi_\Bk \Pi_{-\Bk } \rho(\eta)- \Pi_{-\Bk }\rho(\eta) \chi_{\Bk }\right]
- G^*(\eta) \left[ \rho(\eta) \Pi_{-\Bk } \chi_{ \Bk } - \chi_{\Bk }\, \rho(\eta) \Pi_{-\Bk }\right]\Big \},
}
\end{widetext}
The Hamiltonian in \eqref{drho2} is
\EQ{
H_0 = \frac{1}{2}\sum_\Bk \Big[ \Pi_{\Bk }\Pi_{-\Bk } + k^2 \chi_{\Bk }\chi_{-\Bk }
+ \frac{a'}{a} \left( \chi_{\Bk } \Pi_{-\Bk } + \Pi_{\Bk }\chi_{ -\Bk }\right) \Big].
}
The full density matrix can then be factorised as
\begin{equation}
\rho(\eta) = \prod_{|\Bk |} \otimes \,\rho( \Bk , - \Bk ) + \mathcal{O}(\lambda^3),
\end{equation}
where the modes of different wavelengths only mix at $\mathcal{O}(\lambda^3)$. Under this assumption (remembering to pick out the $+\Bk $ and $-\Bk $ terms from the mode sum in eq.~(\ref{drho2})) we find, dropping the $\Bk $ subscripts, the following master equation for the two-mode density matrix $\rho(\Bk ,-\Bk )$
\begin{widetext}
\EQ{
\frac{d \rho}{d \eta}& = \overbracket{ -i\Big[H_{\text{eff}},\rho\Big] }^{\text{unitary evol.}}
+ \overbracket{\frac{i\lambda^2 }{8 \pi^2 H^2 \eta} \,\text{Re}\, G \,
\Big\{ \left[ \chi ,\lbrace \Pi^\dagger , \rho \rbrace \right] +\left[ \chi^\dagger , \lbrace \Pi , \rho \rbrace \right] \Big\}}^{\text{dissipation/relaxation}} \\
& - \underbracket{\frac{\lambda^2}{16 \pi^2 H^2 \eta^2} \left( \pi + 2\, k \,\eta \, \,\text{Im} \, F\right)
\Big\{ \left[\chi^\dagger,\left[\chi, \rho\right]\, \right] + \left[\chi, \left[\chi^\dagger, \rho\right]\, \right]\Big\}}_{\text{fluctuation/decoherence}}
- \frac{\lambda^2 }{8 \pi^2 H^2 \eta} \,\text{Im} \, G\,
\Big\{ \left[ \chi ,\left[ \Pi^\dagger , \rho \right] \right] +\left[ \chi^\dagger , \left[ \Pi , \rho \right] \right] \Big\} \label{drho1},
}
where the renormalised Hamiltonian $H_{\text{eff}} = H_0 + \delta H$ contains the contributions
\EQ{
\delta H & = - \frac{ k \lambda^2}{4 \pi^2 H^2 \eta}\, \,\text{Re}\,F\, \chi^\dagger \chi, \\
H_0& = k^2 \chi^\dagger \chi + \Pi^\dagger \Pi + \frac{a'}{2 a} \left( \chi \Pi^\dagger+ \Pi \chi^\dagger + \chi^\dagger \Pi+ \Pi^\dagger \chi \right).\label{Hamil}
}
This reveals striking similarities between the dynamics of inflation and quantum Brownian motion as described in sec.~\ref{philosophy}. This provides a deeper insight into the master equation first presented in \cite{Boy1} and will prove important in subsequent analysis. It also paints a richer and more detailed picture than the simple master equation presented in \cite{BLM} (see for instance their eq.~(10)). However, as in \cite{BLM} we too shall see that the master equation is dominated by a simple diffusion term on super-horizon scales. The first term on the right-hand side describes the unitary dynamics (with renormalised Hamiltonian) and the second and third terms correspond respectively to dissipation (i.e.~a friction term) and fluctuations/decoherence of the system due to the environmental noise. The final term, which does not fit the quantum Brownian motion scheme, will be discussed more shortly.
These first three terms are easily identified with corresponding terms in the Caldeira-Leggett master equation of quantum Brownian motion \eqref{mse}. Whilst the similarities are striking, an important difference is that in the cosmological setting the couplings of the various terms are time-dependent.

The interpretation of the dissipation-fluctuation terms becomes more apparent if we write the master equation in terms of matrix elements in the field basis $\rho(\chi_{\Bk }, \tilde{\chi}_{\Bk } ) \equiv \bra{ \chi_{\Bk } } {\rho}(\Bk ,-\Bk ) \ket{ \tilde{\chi}_{\Bk } }$. Given any operator $M$, the action of $\Pi$ is given by
\begin{equation}
\bra{ \chi } {M} {\Pi}^\dagger \ket{ \tilde{\chi} } = i \frac{\partial}{\partial \tilde{\chi}} \bra{ \chi } {M} \ket{ \tilde{\chi} }, \qquad
\bra{ \chi } {M} {\Pi} \ket{ \tilde{\chi} } = i \frac{\partial}{\partial \tilde{\chi}^*} \bra{ \chi } {M} \ket{ \tilde{\chi} }.
\end{equation}
Theses relations follow from the commutation relations $[\chi,\Pi^\dagger] = 1$ and the fact that $\exp(i \chi \Pi^\dagger)$ is the generator of translations. The action of other terms involving $\Pi$ can be inferred from these expressions by appropriate complex conjugation. This gives rise to the following equation for the matrix elements:
\EQ{
\frac{d \rho}{d \eta} & =
- i \Big( k^2 - \frac{k \lambda^2}{4 \pi^2 H^2 \eta} \, \,\text{Re}\,F \Big) \left( |\chi|^2 - |\tilde{\chi}|^2 \right)
+ i \left( \frac{\partial^2 \rho}{\partial \chi \partial \chi^* } - \frac{\partial^2 \rho }{\partial \tilde{\chi} \partial \tilde{\chi}^*} \right) -\frac{a'}{a}\left( \chi \frac{\partial \rho}{\partial \chi} + \chi^*\frac{\partial \rho}{\partial \chi^*} + \tilde{\chi}\frac{\partial\rho}{\partial \tilde{ \chi}} + \tilde{\chi}^* \frac{\partial \rho}{\partial \tilde{\chi}^*} + 2 \rho \right) \\
& + \frac{\lambda^2}{4 \pi^2 H^2 \eta} \,\text{Re} \, G
\left[\left( \chi - \tilde{\chi}\right)\left( \frac{\partial \rho}{\partial \chi} - \frac{\partial \rho}{\partial \tilde{\chi}}\right)+ \left( \chi^* - \tilde{\chi}^* \right) \left( \frac{\partial \rho}{\partial \chi^*}-\frac{\partial \rho}{\partial\tilde{\chi}^*}\right) \right] \\
&- \frac{\lambda^2}{8 \pi^2 H^2 \eta^2} \left( \pi + 2 \, k \eta \, \,\text{Im} \, F\right) \left|\chi - \tilde{\chi}\right|^2 \, \rho\\
& + \frac{i \lambda^2}{4 \pi^2 H^2 \eta} \,\text{Im} \, G
\left[\left( \chi - \tilde{\chi}\right)\left( \frac{\partial \rho}{\partial \chi} + \frac{\partial \rho}{\partial \tilde{\chi}}\right)+ \left( \chi^* - \tilde{\chi}^* \right) \left( \frac{\partial \rho}{\partial \chi^*}+\frac{\partial \rho}{\partial\tilde{\chi}^*}\right) \right], \label{drho}
}
which can be compared with the Brownian motion case in equation (\ref{mse}).
\end{widetext}
One can also use this master equation to study the phase-space of the theory as captured by the Wigner function, cf.~\eqref{wfn},
\EQ{\label{Wig}
&W(\chi,\Pi) = \\
& \frac{1}{\pi^2} \int^\infty_{-\infty} d^2\chi' \rho\left(\chi - \frac{\chi'}{2},\chi + \frac{\chi'}{2}\right)e^{ i2 \chi'_R\Pi_R + i 2 \chi'_I \Pi_I},
}
which provides a quantum analogue of a phase-space probability distribution. By Wigner transforming the master equation (\ref{drho}) we can derive the following equation for $W$, mirroring the case of quantum Brownian motion \eqref{wme},
\EQ{
\frac{\partial W}{d\eta} &= \lbrace{ H_{\text{eff}},W \rbrace} - \frac{\lambda^2}{4 \pi^2 H^2 \eta} \,\text{Re} \, G \left( \partial_\Pi \, (\,\Pi W) + \partial_{\overline{\Pi}}(\bar{\Pi} W) \right) \\
&+ \frac{\lambda^2}{8 \pi^2 H^2 \eta^2}\left( \pi + 2 k \eta \,\text{Im} F \right) \partial_\Pi \partial_{\bar{\Pi}} W \\
&- \frac{\lambda^2}{8 \pi^2 H^2 \eta} \,\text{Im}\, G\left( \partial_\chi \partial_{\bar{\Pi}} W + \partial_{\bar{\chi}} \partial_{\Pi} W \right),\label{FPW}
}
This has the usual terms: decoherence (the diffusion terms involving two-derivatives), the dissipation/friction term (single derivatives) and the renormalisation of the Hamiltonian. Note that in our model, the field $\chi$ has no self-interactions and so there is no analogue of the quantum terms in \eqref{wme}.

\begin{figure}
\centering
\includegraphics[trim={6.8cm 11.5cm 4cm 12cm}]{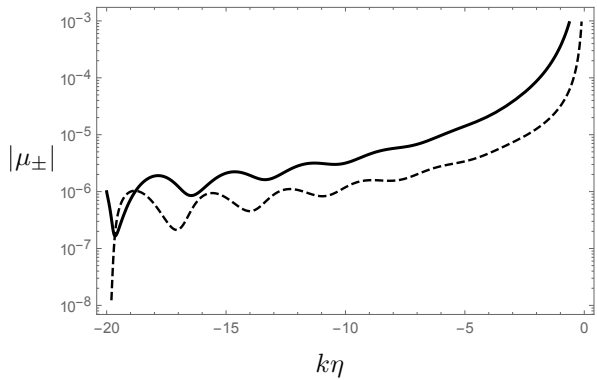}
\caption{\footnotesize The modulus of the eigenvalues $\mu_+$ and $\mu_-$ corresponding to positive (solid) and negative (dashed) diffusion respectively, where we took $\lambda/H=0.1$.}
\label{eigenvalues}
\end{figure}

The diffusion term on the right-hand side can be written in the form $\mathcal{D}_{ij} \partial_i \overline{\partial}_j W$, where the derivatives are defined by $\partial_1 = \partial_\chi$ and $\partial_2 = k \partial_\Pi$, with the diffusion matrix defined by
\begin{equation}
\mathcal{D} =
\frac{\lambda^2}{ 8 \pi^2 H^2 k \eta}\left(
\begin{array}{cc}
0 & - \text{Im}\, G\\
- \text{Im}\, G \qquad& \quad (\pi + 2 k \eta \,\text{Im} F)/(k\eta)
\end{array}
\right).
\end{equation}
This has eigenvalues
\EQ{
\mathcal{\mu}_{\pm}& = \frac{\lambda^2}{16 \pi^2 H^2 k^2 \eta^2 } (\pi + 2 k \eta \,\text{Im} F) \\
& \pm \frac{\lambda^2}{16 \pi^2 H^2 k^2 \eta^2 }\sqrt{\left( \pi + 2 k \eta \,\text{Im} F\right)^2 +4 \left( k \eta \text{Im}\, G \right)^2}.
}
We see immediately that the eigenvalue $\mu_-<0$. In fact, the plot in Fig.~\ref{eigenvalues} shows that these eigenvalues remain comparable in magnitude, so that positive and negative diffusion, at first sight appear to be of equal importance. This can also be seen from the asymptotic form of $F$ and $G$ for $|\eta k| \ll 1$
\EQ{
F(\eta) & \simeq - \frac{1}{k \eta}\Big( 1 - \log \frac{\eta}{\eta_0}\Big), \\
G(\eta) & \simeq \frac{i}{2 k \eta} + \frac{1}{3}\Big(1 - \log \frac{\eta}{\eta_0} \Big), \label{asympGpm}
}
which reveal that $\mu_{\pm}$ scale as $\pm1/\eta^2$ on super-horizon scales. This ``anti-diffusion" has no classically meaningful analogue, and can be traced to the $\text{Im}\, G$ term in master equation (\ref{drho}).

The other violation of positivity comes from the behaviour of the coefficients in the master equation, which exhibit oscillatory behaviour, fluctuating sign at early times. Similar behaviour is seen for coefficients of the master equation in flat space \cite{Boy1}, where the oscillations are due to the non-local environment correlations corresponding to memory effects. These effects are usually dealt with by temporal coarse-graining in the form of a kind of rotating wave approximation. This approximation is key to writing the master equation in Lindblad, and hence manifestly positivity preserving form.

At this stage, our situation may seem a little perilous: we have a master equation which is not in Lindblad form and exhibits several features, e.g.~anti-diffusion and oscillatory coefficients, which might jeopardise the positivity of the density matrix. In fact, although the unitary evolution of the total state ensures positivity of the full density matrix, the series of simplifying approximations outlined in sec.~\ref{philosophy} and carried out explicitly in sec.~\ref{masterEq} provide no guarantee that the spectrum of the reduced density matrix remains positive. Whilst Hermiticity and normalisation follow trivially, the most important property -- that $\rho(\eta)$ be positive definite -- remains to be established. Indeed, even the Caldeira-Leggett model can lead to a small violation of positivity \cite{H,BP}. In fact, we shall see the operator form of the master equation and the oscillatory behaviour of the coefficients mean that the master equation manifestly violates positivity on sub-horizon scales.

However the master equation alone does not tell the full story. The real question is whether physically relevant solutions of the master equation maintain positivity as a function of time. This shall be the subject of the following sections.

\subsection{Solving the master equation}\label{SolSec}

In accordance with our previous boundary conditions, we shall take Bunch Davis initial conditions which imply a Gaussian ansatz for the density matrix
\begin{equation}\label{Gaussian}
\rho(\chi, \tilde{\chi})= \mathcal{N}\exp\left\{ - \Omega\chi^2 - \Omega^*\tilde{\chi}^{2} - \frac{1}{2}\xi|\chi - \tilde{\chi}|^2\right\},
\end{equation}
and initial conditions
\begin{equation}
\Omega (\eta_0) = 1, \qquad \xi(\eta_0)=0, \qquad \mathcal{N}(\eta_0) = \frac{2 \Omega_R}{\pi},
\end{equation}
where $\Omega = \Omega_R + i \Omega_I$ is complex and $\xi$ is real, ensuring the Hermiticity of the density matrix. Notice that although we use the notation $\Omega_R$ for the Gaussian width, it does \textit{not} describe a simple isolated squeezed state, and, in particular, is not a simple expression which can be written in terms of squeezing parameters $r_k$ and $\varphi_k$ of \textit{e.g.} \cite{Lim,VM}. On the contrary, $\Omega$ contains $\mathcal{O}(\lambda^2)$ corrections and in general must be solved numerically, as we shall see below.

Substituting the ansatz (\ref{Gaussian}) into eq. (\ref{drho}), and comparing the coefficients of $|\chi|^2$, $|\tilde{\chi}|^2$, $\chi \tilde{\chi}^*$ and $\tilde{\chi} \tilde{\chi}$ gives the evolution equations for $\Omega$ and $\xi$ and $\mathcal{N}$:
\begin{widetext}
\begin{align}
\frac{d \Omega}{d\eta }& = i k^2 +\frac{2}{ \eta}\Omega - i \Omega_R \xi - i \Omega^2 - \frac{ i \lambda^2}{4 \pi^2 H^2 \eta} \Big( k \,\text{Re}\, F - \,\text{Im} \left( G \Omega \right) \Big), \label{dOmega} \\[5pt]
\frac{d \xi}{d \eta} & = 2 \xi \Big( \Omega_I + \frac{1}{\eta} \Big) + \frac{\lambda^2}{4 \pi^2 H^2 \eta^2}\Big( \pi + 2 \eta k\,\text{Im}\, F + 2 \eta \,\text{Re}\,\left( G \Omega\right) + 2 \eta \,\text{Re}\, G \xi\Big),\label{dzeta} \\[5pt]
\frac{d \mathcal{N}}{d\eta}& = 2 \,\mathcal{N}\Big( \Omega_I + \frac{1}{\eta} \Big).\label{dN}
\end{align}
\end{widetext}
Note that $F$ and $G$ are in fact dimensionless, and functions only of the dimensional combination $k \eta$, as can be seen from eqs.~(\ref{F}) and (\ref{G}).

We immediately recognise that $\xi$ governs decoherence since it controls the suppression of the off-diagonal elements of $\rho$, corresponding to the decay of quantum interference. Notice also that by taking the real part of eq.~(\ref{dOmega}) and using the normalisation condition (\ref{dN}) we get
\begin{equation}\label{normalisation}
\frac{1}{\Omega_R}\frac{d \Omega_R}{d\eta}=\frac{2}{\eta}+2\Omega_I= \frac{1}{\mathcal{N}} \frac{d \mathcal{N}}{d\eta},
\end{equation}
whose solutions give the expected normalisation condition $\mathcal{N} = 2 \Omega_R/\pi$ for a (complex) Gaussian, providing a consistency check for the evolution equations.

Using these equations, and the asymptotic behaviour for $F$ and $G$ given in eq.~(\ref{asympGpm}) we find the following super-horizon behaviour for $\Omega_R$, $\Omega_I$ and $\xi$ in the $|k \eta| \ll 1 $ limit:
\begin{align}
\Omega_R & = k^3 \eta^2 \exp \left[ \frac{\lambda^2}{12 \pi^2 H^2} \Big( \log ^2 \frac{\eta}{\eta_0} - \frac{4}{3} \log \frac{\eta}{\eta_0}\Big)\right] \nonumber \\
& +O\left( \eta^3\right) ,\label{dOR}\\[5pt]
\Omega_I &=\frac{ \lambda^2}{12 \pi^2 H^2}\frac{1}{\eta}\left(\log \frac{\eta}{\eta_0}-\frac{2}{3}\right) - k^2 \eta+O\left(\eta^3\right), \label{dOI} \\
\xi &= - \frac{ \lambda^2}{12 \pi^2 H^2} \frac{1}{\eta}+ O\left(\eta\right). \label{Dzeta}
\end{align}
These expressions will be used throughout the remainder of this paper to study the super-horizon behaviour of fluctuations.

It is noteworthy that the expression for $\Omega_R$ in \eqref{dOR} has the form of a re-summed quantity in perturbation theory because it is $\Omega_R$ that determines the all-important power spectrum that is a key observable for the CMB. This re-summed expression was first derived in \cite{Boyanovsky:2015jen} and
in order to make contact with that analysis, let us focus in more detail on the equation for $\Omega_I$. Including the bare mass term for $\chi$ and the UV divergence, this equation takes the form
\EQ{
\frac{d\Omega_I}{d\eta}&=k^2+\frac{{\cal M}_\chi^2}{H^2\eta^2}+\frac{\lambda^2}{4\pi^2H^2\eta^2}\log\frac\varepsilon{-\eta_0}\\ &+\frac2\eta\Omega_I+\frac{\lambda^2}{4\pi^2H^2\eta^2}\Big(1-\log\frac\eta{\eta_0}\Big)+\cdots.
\label{omi}
}
Note that this expression makes clear that the subtraction point we chose to be $\eta_0$ is in fact arbitrary. Let us set the renormalized mass to zero but include a specific finite counter-term:
\EQ{
{\cal M}_\chi^2=-\frac{\lambda^2}{4\pi^2}\log\frac\varepsilon{-\eta_0}-\frac{\lambda^2}{6\pi^2}.
}
Solving \eqref{omi} and then integrating \eqref{normalisation} to find $\Omega_R$ (and fixing the constant of integration appropriately), we find
\EQ{
\Omega_R&=k^3\eta^2\exp\Big[\frac{\lambda^2}{12\pi^2 H^2}\Big(\log^2(-k\eta)\\ &\qquad\qquad-2\log(-k\eta)\log(-k\eta_0)\Big)\Big].
\label{kek}
}
This quantity gives directly the power spectrum of the scalar perturbations:
\begin{widetext}
\EQ{
{\cal P}(k,\eta)=\frac{k^3H^2\eta^2}{2\pi^2}\cdot\langle\chi_\Bk\chi_\Bk^\dagger\rangle
=\frac{k^3H^2\eta^2}{(2\pi)^2}\cdot \frac1{\Omega_R}
=\frac{H^2}{(2\pi)^2}\exp\Big[\frac{\lambda^2}{12\pi^2 H^2}\Big(2\log(-k\eta_0)\log(-k\eta)-\log^2(-k\eta)\Big)\Big].
\label{kek2}
}
\end{widetext}
This is the result quoted in \cite{Boyanovsky:2015jen} for the correction to the power spectrum. Note that the coefficient of the $\log(-k\eta)$ in the exponent can be altered by changing the finite part of the counter-term while the $\log^2(-k\eta)$ term is universal.


\subsection{Positivity on super horizon scales and the nature of decoherence}\label{positivity}

\begin{figure}[ht]
\begin{center}
\includegraphics{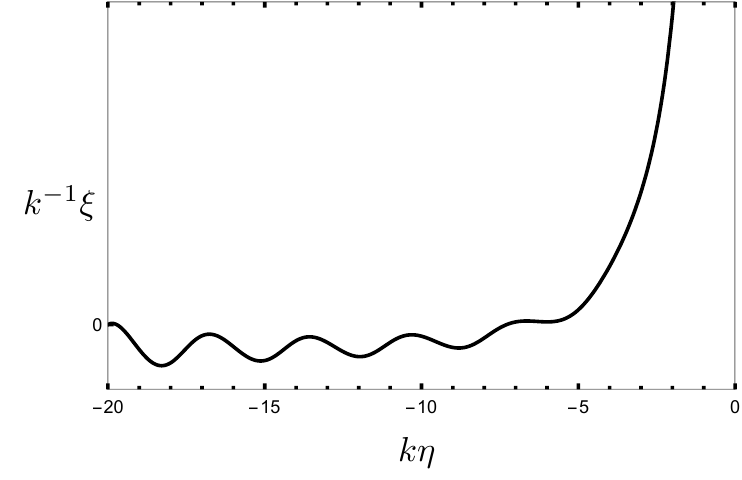}
\caption{\footnotesize The decoherence parameter $\xi$ as a function of $k\eta = -k/aH$.}
\label{zeta}
\end{center}
\end{figure}

Positivity is guaranteed if, and only if, $\rho$ has non-negative eigenvalues. To find these, we note that the density matrix (elements) factorise into a product of two identical Gaussians of the real and imaginary parts of $\chi$, i.e.,
\begin{equation}\label{matrix_elt_1}
\rho(\chi, \tilde{\chi}) = \rho_I(\chi_I , \tilde{\chi}_I)\rho_R(\chi_R , \tilde{\chi}_R),
\end{equation}
where
\EQ{\label{rhoR}
&\rho_R(\chi_R , \tilde{\chi}_R) \\
& = \sqrt{\frac{2 \Omega_R}{\pi}} \exp\Big[ - \Omega \chi_R^2 - \Omega^* \tilde{\chi}_R^2 - \frac{\xi}{2} ( \chi_R - \tilde{\chi}_R)^2 \Big],
}
with an identical expression for $\rho(\chi_I , \tilde{\chi}_I)$ given by replacing $\chi_R \rightarrow \chi_I$. The eigenstates of $\rho$ are then simply tensor products $\rho_I$ and $\rho_R$ eigenstates, so that positivity of $\rho$ is then equivalent to positivity of $\rho_I$ and $\rho_R$. Following \cite{JZ}, we can write $\rho_R$ in diagonal form as
\begin{equation}
\rho_R = \sum_n p_n \ket{\varphi_n} \bra{\varphi_n}.
\end{equation}
The eigenstates $\ket{\varphi_n}$ and their eigenvalues $p_n$ satisfy (dropping the $R$ subscript on $\chi$)
\begin{equation}\label{eigenEQ}
p_n \, \varphi_n(\chi) = \int d \chi' \rho_R(\chi ,\chi') \varphi_n(\chi'),
\end{equation}
where $\varphi_n(\chi) \equiv \braket{\chi | \varphi_n}$ gives the position representation of the $\left\{\ket{ \varphi_n }\right\}$. The Gaussian form of $\rho_R$ means that $\varphi_n(x)$ take the same form as harmonic oscillator wave functions, i.e.,
\begin{equation}
\varphi_n(\chi) = N H_n(\alpha \chi) \exp(-\beta \chi^2),
\end{equation}
where $H_n$ are the Hermite polynomials, $N$ is a normalisation factor, and $\alpha$ and $\beta$ are functions to be determined. Substituting this into (\ref{eigenEQ}) and using the integral identity in \cite{GR} we find
\EQ{\label{alphabeta}
\alpha^2 &=2 \left[\Omega_R (\Omega_R + \xi)\right]^{1/2}, \\
\beta &=i \Omega_I + \sqrt{\Omega_R (\Omega_R + \xi)}.
}
We can also read off
\begin{equation}\label{pn}
p_n = p_0 (1-p_0)^n,
\end{equation}
where
\begin{align}
p_0 & = \frac{2 }{1 + \sqrt{1+ \xi/\Omega_R} }.\label{p0}
\end{align}
Positivity requires $0 \leq p_0 \leq 1$, which, assuming $\Omega_R>0$, holds if and only if $\xi >0$. From the numerical solution for $\xi$ in Fig.~\ref{zeta}, one can see that at early times positivity is violated, but that $\xi$ becomes positive as the modes exit the horizon.
%
%
%


This can be understood by examining the equation (\ref{dzeta}) for $\xi$, which can be re-written in the form
\EQ{\label{dzetadu}
\frac{d \xi}{d\eta} & = \Big[ \frac{ \lambda^2}{2 \pi^2 H^2 \eta} \,\text{Re}\, G +
\frac{1}{\Omega_R} \frac{d \Omega_R}{d\eta} \Big] \xi \\
& + \frac{\lambda^2}{4 \pi^2 H^2 \eta^2} \left[ \pi + 2 k \eta \,\text{Im}\,F + 2 \eta \,\text{Re} \,(G \Omega) \right].
}
One notes that on super-horizon scales ($|k \eta| \ll 1$) the non-local (memory) effects in the first square braket $[...]\xi$ corresponding to $\text{Re}(G)$ become subdominant, giving way to the leading behaviour from $\Omega_R^{-1}{d \Omega_R}/{d \eta}\simeq 2/\eta$. Similarly the ``forcing term" in the second square-braket is dominated by the first term (which is ultimately traced to the local contribution in the master equation) with the non-local effects from $\text{Im} F$ and $\text{Re}(G\Omega)$ becoming negligible. Thus on super-horizon scales we find
\begin{equation}\label{SqzDec}
\frac{d \xi}{d\eta} \simeq \underbracket{\,\, \frac{2}{\eta}\,\,\xi}_{\text{squeezing}} + \underbracket{\,\,\frac{ \lambda^2}{4 \pi H^2 \eta^2}\,\,}_{\text{interaction}},
\end{equation}
The second term in (\ref{SqzDec}) drives the positive growth of $\xi$. It originates from the local term in the first line of (\ref{drho2}) and is associated to the decoherence term in (\ref{drho}). Equivalently, the relevant super-horizon dynamics in the evolution equation for $\xi$ can be reproduced by replacing the environment correlators according to
\begin{equation}
K(k,\eta,\eta') \longrightarrow \frac{1}{8 \pi^2}\delta(\eta'-\eta),
\end{equation}
back in eq.~(\ref{rhoX}); in other words, long time environmental correlations become unimportant for decoherence at late times, which is driven only by, what will become in the stochastic interpretation, white noise. This is a common feature of many systems, where they exhibit essentially memoryless (Markov) behaviour at late times, when long time environmental correlations become negligible. This is precisely the case for the flat space master equation considered in \cite{Boy1}, where initially oscillatory coefficients in the master equation, induced by short-time environment correlations lead to positivity on long-times. Hence, we see that the oscillatory behaviour in the interaction terms associated to the kernels $F$ and $G$, which can be traced to non-local/memory effects, leads to an initial violation of positivity. However, at late times, this gives way rather elegantly to the dominance of effects driven by local terms and their white noise leading to
\begin{equation}
\xi \simeq \frac{\lambda^2}{12 \pi H^2} \frac{1}{|\eta|}\quad \text{for} \quad |k \eta| \ll1.
\end{equation}

It should be noted in passing however, that by contrast, the object $\Omega_R$ is dominated by \textit{non-local} contributions to the master equation on super-horizon scales. In this sense, the power spectrum receives corrections due to memory effects -- see eq.~(\ref{kek2}). We mention this purely to point out that although $\xi$ becomes dominated by local effects, there remain other physically interesting quantities which \textit{are} dominated by memory effects in the master equation. Indeed, the resummation of non-local contributions and their secular log terms is one of the key results in Boyanovsky's paper \cite{Boyanovsky:2015jen}.

The growth in $\xi$ can be traced back to the fact the interaction strength is effectively time-dependent and grows with time, as can be seen from eq.~(\ref{interaction}), where one could in principle absorb the time-dependence into the coupling constant by taking $\lambda^2 \rightarrow \lambda^2(\eta) \equiv \lambda^2/ H^2 \eta^2$. It is therefore interesting to ask how generic this growth in $\xi$ is, and what happens when one looks at other interactions, e.g.~marginal couplings such as $\mathcal{H}_{int} = \sqrt{-g} \phi^2 \varphi^2 $ as well as other relevant couplings to see how model-dependent the scale-dependence of decoherence is. In addition this might save us from some of the problems associated with having a coupling constant that grows in time, thus invalidating perturbativity at late times. We leave such investigations for future work.

At this point we also highlight an important subtlety in the evolution equation for $\xi$. Let us consider the first term in eq.~(\ref{SqzDec}) of the form $2/\eta$ which corresponding to squeezing. This can be traced to the $\Omega_R^{-1} \Omega_R'$ term in (\ref{dzetadu}) or equivalently to the $a'/a$ term in the Hamiltonian (\ref{Hamil}). Notice that this term gives a negative contribution to $d \xi/d\eta$. At first sight, this makes things look as though the squeezing is slowing the increase in $\xi$ and therefore resisting the suppression of off-diagonal elements. This would certainly run counter to the party line that squeezing makes a state more classical. However, such reasoning is misleading as we now explain.

\begin{figure}
\centering
\includegraphics[scale=0.47]{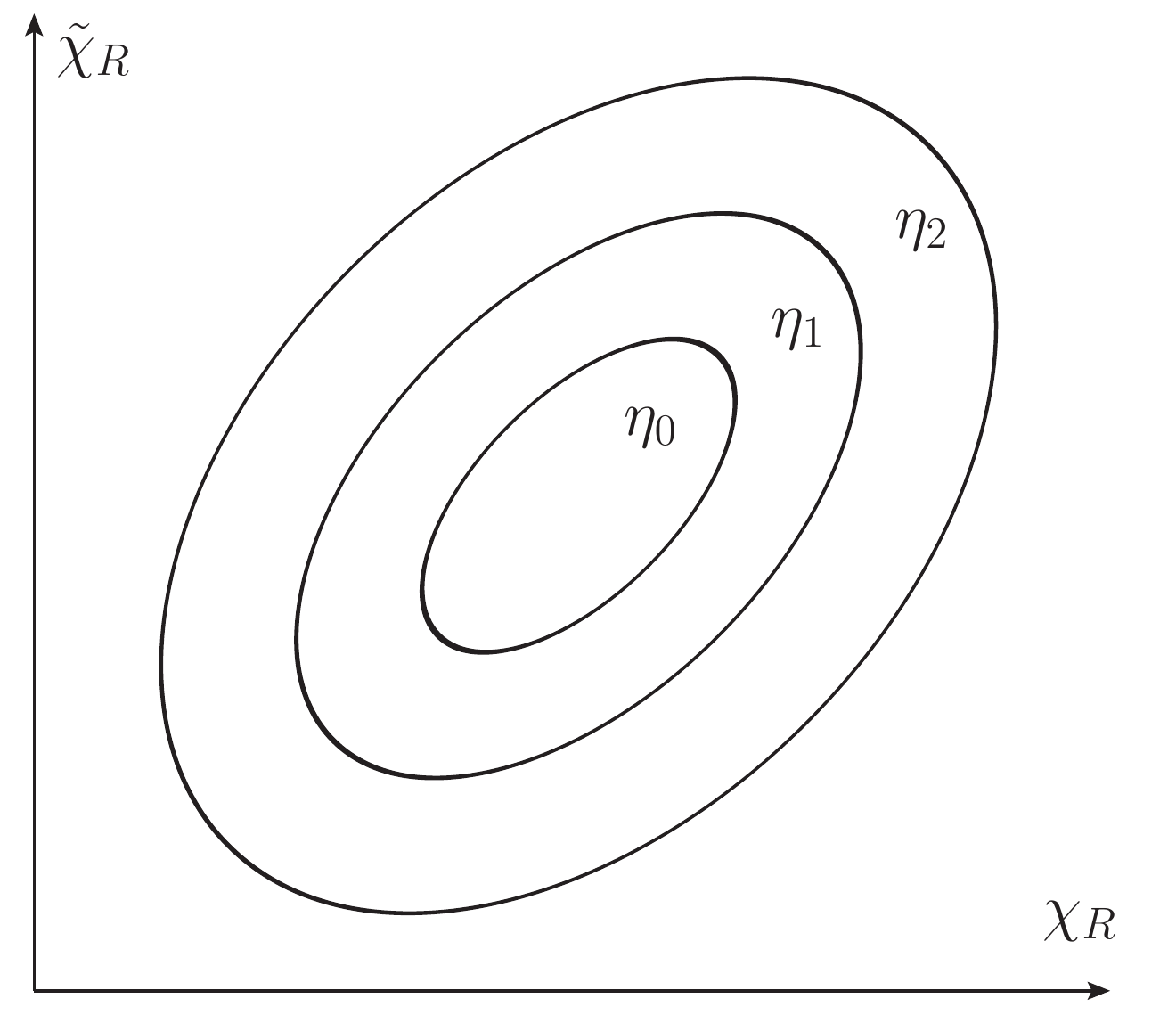}
\caption{\footnotesize{The contours of $|\rho| =const $ at times $\eta_0 <\eta_1 < \eta_3$ in the free case $\lambda =0$. The ellipse as a greater extent in the ``off-diagonal direction" due to overall scaling as $\Omega_R \rightarrow 0$ but does not change shape.} }
\label{RhoEllipse}
\end{figure}

This can be seen most clearly in the free case with $\lambda =0$. Explicitly, the evolution equations become
\EQ{
\frac{d \xi}{d\eta} & = \frac{1}{\Omega_R} \frac{d \Omega_R}{d\eta}\xi, \\
\frac{d \Omega}{d\eta} & = ik^2 + \frac{2 \Omega}{\eta} - i \Omega_R \xi - i \Omega^2.
}
One finds that the solutions satisfy $\xi \propto \Omega_R$. Since $\Omega_R(\eta_0)=1$, solutions are given by $\xi = \xi_0 \Omega_R$. Again, this makes it very tempting to suggest $\Omega_R$ drives $\xi$ to zero. However, $\rho$ takes the form
\begin{equation}
\rho_\text{free}(\chi, \tilde{\chi}) = \frac{2 \Omega_R}{\pi} \exp\left(- \Omega_R \left[ |\chi| + |\tilde{\chi}| + \frac{\xi_0}{2}|\chi - \tilde{\chi}|^2 \right]\right).
\end{equation}
From this we see that $\Omega_R$ does decrease the size of $\xi$, and therefore \textit{increases} the off-diagonal extent of the density matrix, but only in the sense that as time passes, it rescales the entire ellipse
\begin{equation}
|\chi| + |\tilde{\chi}| + \frac{\xi_0}{2}|\chi - \tilde{\chi}|^2 = \Omega_R^{-1},
\end{equation}
corresponding to surfaces of $|\rho(\chi,\tilde{\chi})| = \text{constant}$. This is illustrated in the figure \ref{RhoEllipse}. This clearly shows how although $\Omega_R \rightarrow 0$ does cause $\xi$ to shrink (as off-diagonal elements grow in size), the relative size of diagonal and off-diagonal elements stays the same, as can be seen from the fact the contours retain their shape. So in actual fact, squeezing only decreases $\xi$ in a trivial sense in that $\Omega_R$ scales the size of all density matrix elements and so contrary to initial impressions, squeezing does not, in any real sense, resist the onset of decoherence. Explicitly, in the free case, the Wigner area and entanglement entropy depend only on the ratio $\xi/\Omega_R$ which is constant as explained above. Therefore these quantities remain constant -- consistent with the idea that squeezing cannot make the state any ``less-decohered" since it is associated to unitary evolution alone. Similarly the purity $\text{Tr}[\rho^2 ]$ remains constant due to the unitary evolution of the von Neumann equation. When interactions are switched on, we again have the $\Omega_R^{-1} \Omega_R'$ term, but the situation is exactly the same -- this acts only as an overall rescaling and comes from the unitary dynamics.
\subsection{Langevin equation}\label{LangevinEQ}

We now show how it is possible to derive a classical Langevin equation which reproduces the behaviour of the density matrix $\rho(\chi, \tilde{\chi})$ on super-Hubble scales. By this, we mean that, mirroring the case of quantum Brownian motion, the master equation for the Wigner function $W(\chi,\Pi)$ can be interpreted as a Fokker-Planck equation associated to an auxiliary classical stochastic system whose dynamics is described by a Langevin equation. The interpretation is then to go beyond the mere idea that the classical system is auxiliary to the idea that it describes the classical (stochastic) trajectories that have emerged from the quantum system.

To study the super-horizon behaviour of solutions, we return to equations (\ref{dOmega}) and (\ref{dzeta}). From the asymptotic behaviour of $G$ and $F$, (\ref{asympGpm}) and eqs.~(\ref{dOR})-(\ref{Dzeta}), we see that on super-horizon scales the contribution to solutions from $\,\text{Im} \, G$ is sub-leading because it is multiplied by $\Omega_R$ which goes to 0 like $\eta^2$. Thus, at the level of the solutions, squeezing ensures that the anti-diffusion term turns out to be negligible. The late-time behaviour is therefore captured by the Fokker-Planck-like equation
\begin{align}
\frac{\partial W}{d \eta} &=\lbrace{ H_{\text{eff}},W \rbrace} - \frac{\lambda^2}{4 \pi^2 H^2 \eta} \,\text{Re} \, G \left( \partial_\Pi \, (\,\Pi W) + \partial_{\overline{\Pi}}(\bar{\Pi} W) \right) \nonumber \\
&+ \frac{\lambda^2}{8 \pi^2 H^2 \eta^2}\left( \pi + 2 k \eta\,\text{Im} \,F\right) \partial_\Pi \partial_{\bar{\Pi}} W,
\label{efp}
\end{align}
valid in the region $k \eta \to0$,
which, by virtue of the late-time behaviour of $F$, has a positive diffusion matrix.

We now go the extra step and argue that from the perspective of the super-Hubble physics, a classical stochastic dynamics emerges from the underlying quantum system whose Fokker-Planck equation is \eqref{efp}. The individual trajectories of the classical system are solutions of the Langevin equations for the classical variable $\chi$:
\EQ{
\chi' & = \Pi - \frac{\chi}{\eta}, \\
\Pi'& = \Big(\frac{1}{\eta} +\frac{k\lambda^2}{4 \pi^2 H^2 \eta} \,\text{Re} \, G \Big) \Pi \\
&- \Big( k^2+\frac{{\cal M}_R^2}{H^2\eta^2} - \frac{k \lambda^2 }{4 \pi^2 H^2 \eta} \,\text{Re} \, F\Big) + \sigma,
}
where we have allowed for a renormalised mass. The above can be rendered as a single equation for $\chi$:
\EQ{
&\chi'' - \frac{k\lambda^2}{4 \pi^2 H^2 \eta } \,\text{Re}\, G\, \chi' + \Big( k^2+\frac{{\cal M}_R^2}{H^2\eta^2}\\ & - \frac{2}{\eta^2} - \frac{k \lambda^2}{4 \pi^2 H^2 \eta} \,\text{Re} \, F - \frac{k\lambda^2}{4 \pi^2 H^2 \eta^2} \,\text{Re}\, G\Big)\chi= \sigma,
\label{Lang}
}
where the white noise $\sigma$ satisfies
\begin{equation}
\mathbb E(\sigma(\eta)) =0, \qquad \mathbb E(\sigma(\eta) \sigma(\eta')) =\frac{ \lambda^2}{8\pi H^2 \eta^2 } \delta(\eta-\eta').
\end{equation}
This provides a set of stochastic equations capable of reproducing the same super-Hubble dynamics of the quantum master equation in the sense that we have explained in sec.~\ref{philosophy}.
It should be noted that this is quite different from the kind of stochastic equations encountered in the usual stochastic inflation paradigm \cite{Star1, SY} where Fokker-Planck equations emerge even in the absence of interactions, where sub-Hubble modes are interpreted as noise. By contrast, our noise arises from a genuine coupling of different Fourier modes between $\chi$ and $\psi$ so that there is no noise in the $\lambda \rightarrow 0$ limit.

It is interesting that the alternative formalism for dealing with open quantum systems, the influence action approach leads directly to a Langevin equation. For the scalar perturbations, the influence-action Langevin equation was derived in \cite{Boyanovsky:2015jen}. This equation appears to be very different from the one we have written down here in that it features a memory integral and also Gaussian coloured noise. In fact, one can easily show that if one substitutes the zeroth order expression for $\chi(\eta')$ from \eqref{pop} in the memory integral on the right-hand side of eq.~(3.41) of
\cite{Boyanovsky:2015jen} one gets precisely the homogeneous terms in our Langevin equation \eqref{Lang}. Therefore, the difference between the Langevin equations is simply the fact that our equation \eqref{Lang} has only the local (white) noise component of the coloured noise term of the Langevin equation in \cite{Boyanovsky:2015jen}. But we have argued that this is a good approximation for super-horizon modes.

We now solve the equations (\ref{dOmega})-(\ref{dN}) numerically and examine the evolution of various quantities which characterise the emergence of classical stochastic behaviour as modes exit the horizon.

Since from now on we shall work mode-by-mode, we shall work in units where $k=1$ for the remainder of the paper, which greatly simplifies notation by effectively removing any $k$ from most expressions. Dimensionful expressions can always be restored by appropriate insertions of $k$. Throughout what follows we shall take the indicative values
\begin{equation}
\frac{\lambda}{H} = 0.1, \qquad k\eta_0 (= \eta_0) = - 20.
\end{equation}
Obviously the initial value for $|k \eta_0|$ is unrealistically small, but the results are rather insensitive to its actual value.

\subsection{Entropy}\label{EntropySec}
%

\begin{figure}[ht]
\begin{center}
\includegraphics{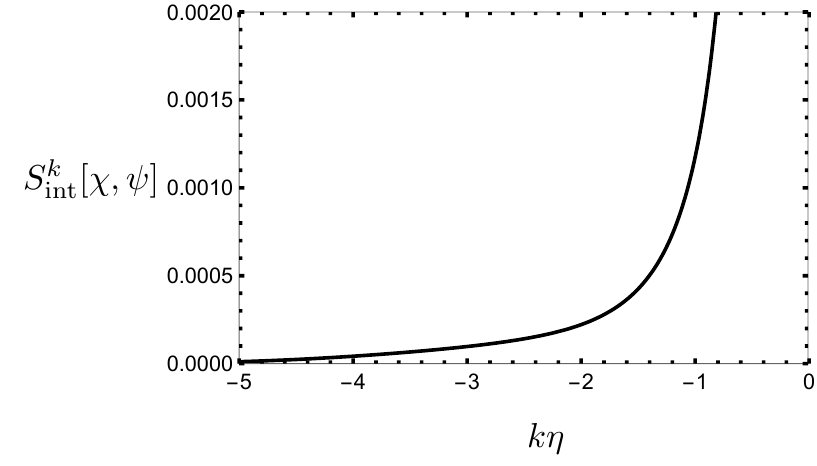}
\end{center}
\caption{\footnotesize The entanglement entropy between the fields $\chi$ and $\psi$ as a function of $k\eta$. Note that the mode exits the horizon at $k\eta=-1$.}
\label{entropy}
\end{figure}

One of the most natural ways to capture the effects of decoherence is to compute the entropy of an open system. In the present context this is the
entanglement entropy between the field $\chi$ (or rather its modes labelled by $k$) and the environment field $\psi$, $S^k_\text{ent}[\chi,\psi]$. This entanglement entropy is, of course, identified with the von Neumann entropy of the reduced density matrix $\rho$ of the $\chi$ field $S_\text{vN}=-\text{Tr}(\rho\log\rho)$.
The growth of entropy (Fig.~\ref{entropy}) provides an important test for the emergence of classicality and the process of decoherence as the system evolves from an initially pure to a mixed state.

Given that the state factorizes $\rho=\rho_R\rho_I$, as in \eqref{matrix_elt_1}, the total entropy is the sum of identical contributions from $\rho_R$ and $\rho_I$, i.e.
\begin{equation}
S^k_\text{ent}[\chi,\psi]\equiv S_\text{vN}(\rho) = S(\rho_I ) + S(\rho_R ) = 2 S(\rho_R).
\end{equation}
where
\begin{align}
S(\rho_R) & = - \sum_{n=0}^\infty p_n \log p_n ,
\end{align}
with $p_n$ given by eqs.~(\ref{pn}) and (\ref{p0}). We find
\EQ{
S(\rho_R)& = - \log \frac{2 }{\sqrt{1+ \xi/\Omega_R} + 1} \\
& -\frac{\sqrt{1 + \xi/\Omega_R} - 1 }{2}
\log \frac{ \sqrt{1+ \xi/\Omega_R} - 1 }{\sqrt{1 + \xi/\Omega_R} + 1}, \label{SrhoR}
}
with the corresponding plot shown in Fig.~\ref{entropy}.

Notice that $S(\rho)$ depends on the ratio $\xi/\Omega_R$, so that the emergence of classical stochastic behaviour (as the state becomes more mixed), as characterised by an increase in entropy, depends on the relative strength of squeezing and decoherence. In the approximation of \cite{KP,KPS2000,Kiefer:2006je}, $\xi$ was assumed to be constant, and so this ratio grows as $\xi/\Omega_R \sim \eta^{-2}$, whereas in reality, our analysis reveals it scales as $\eta^{-3}$ due to the time-dependence of $\xi \sim \eta^{-1}$. Hence our analysis reveals that the growth in entropy is even faster. The same is true for the Wigner area discussed below. This demonstrates the importance of a proper analysis of the evolution of the density matrix in realistic systems.

\subsection{Wigner function and phase space}

\begin{figure}[ht]
\begin{center}
\includegraphics{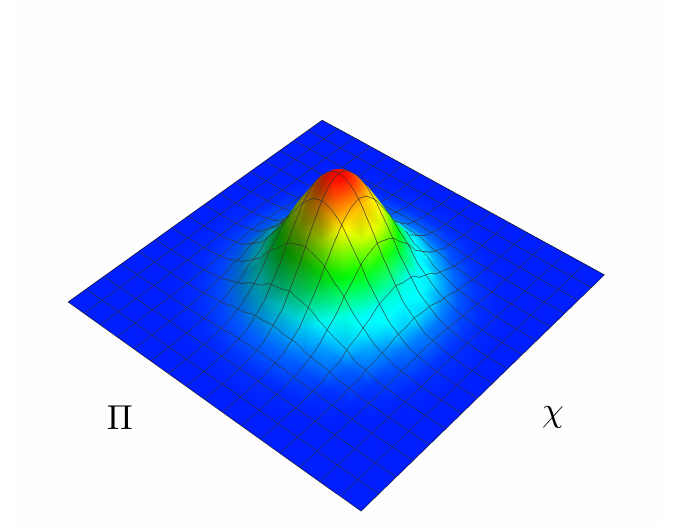}
\caption{\footnotesize Wigner function $W(\chi_R,\Pi_R)$ at $\eta=\eta_0$.}
\label{W1}
\end{center}
\end{figure}

\begin{figure}[ht]
\begin{center}
\includegraphics[scale=0.4]{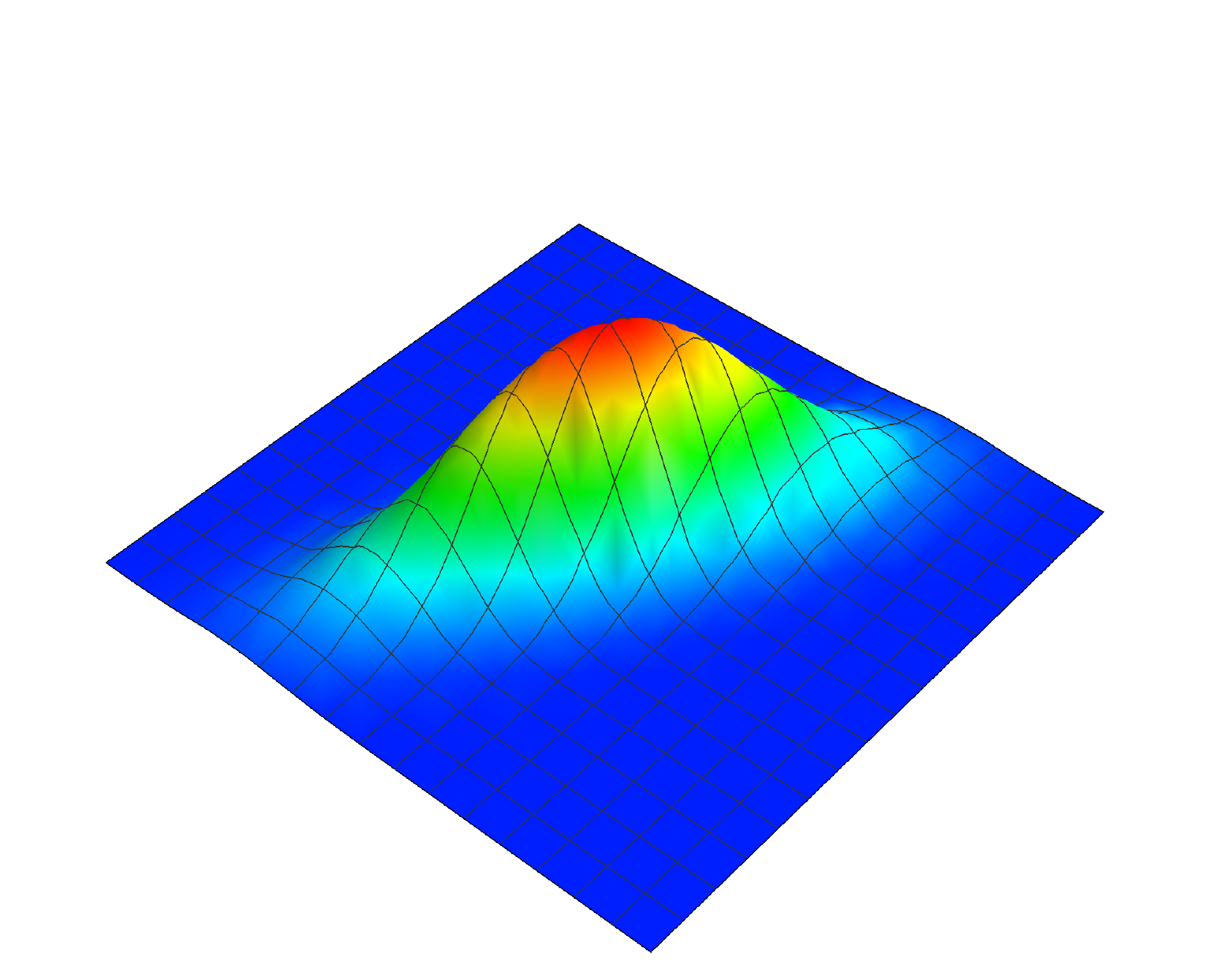}
\end{center}
\caption{\footnotesize Wigner function at $k \eta=-0.8$.}
\label{W2}
\end{figure}

\begin{figure}[t]
\begin{center}
\includegraphics{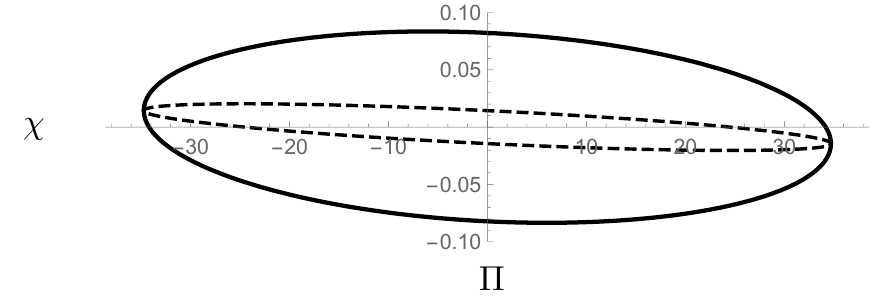}
\caption{\footnotesize Squeezed Wigner ellipse on super-horizon scales at $k\eta=-0.02$ in the interacting (solid) and free (dashed) cases. }
\label{twoellipse}
\end{center}
\end{figure}

\begin{figure}
\begin{center}
\includegraphics{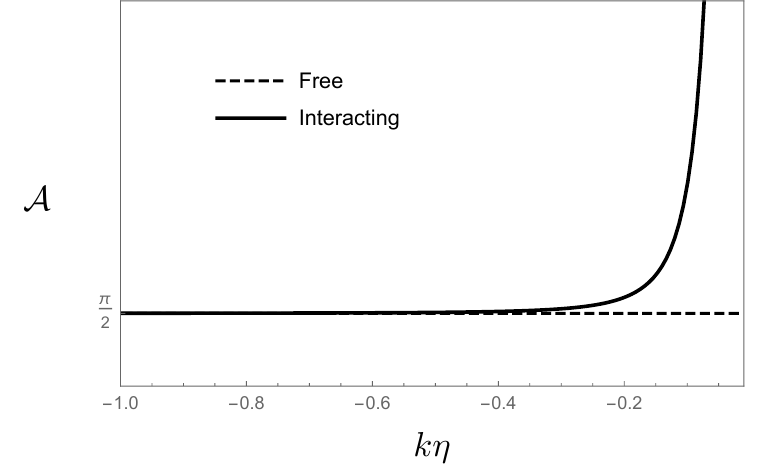}
\caption{\footnotesize Area of Wigner ellipse associated to $W(\chi_R,\Pi_R)$.}
\label{area}
\end{center}
\end{figure}

\begin{figure}[t]
\begin{center}
\includegraphics{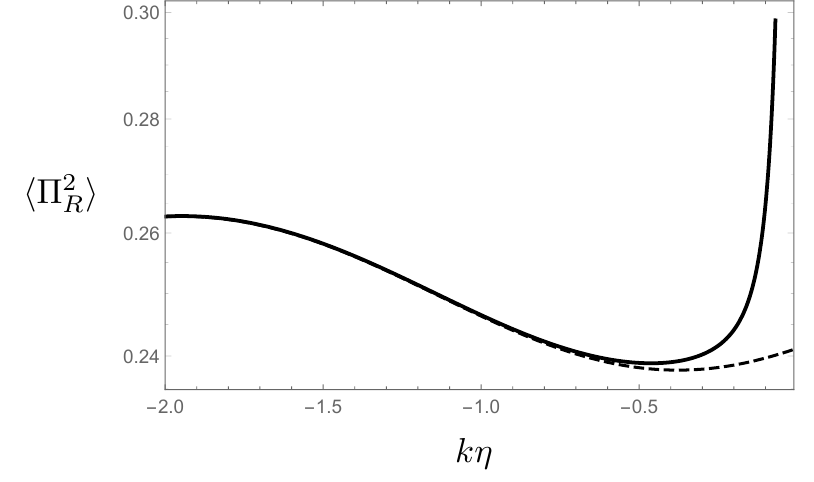}
\caption{\footnotesize Momentum correlator $\braket{\Pi_R^2}$ associated to $W(\Pi_R,\chi_R)$ in interacting case (solid) and free case (dashed).}
\label{PiPi}
\end{center}
\end{figure}

The trade-off between squeezing and decoherence can be better understood by exploring the phase space portrait of the state as provided by the Wigner function. Explicitly, by performing the integral in eq.~(\ref{Wig}) with the Gaussian ansatz (\ref{Gaussian}), we get
\EQ{
W(\chi,\Pi) =\frac{4}{\pi^2} \frac{\Omega _R}{\xi +\Omega _R}
\exp \Big[-\frac{2 \left|\Pi+ \Omega _I \chi \right|^2}{\Omega _R + \xi } - 2 \Omega _R \left|\chi \right| ^2\Big].
}
Notice that this factorises into two identical Wigner functions for $W(\chi_R,\Pi_R)$ and $W(\chi_I,\Pi_I)$. The evolution of $W(\chi_R,\Pi_R)$ is plotted in figures \ref{W1} and \ref{W2}.
Using this, one can compute the spread in the ``position" and ``momentum" of the field as captured by the correlators
\begin{equation}\label{correlators}
\braket{\chi_R^2} = \frac{1}{4\Omega_R}, \qquad \braket{\Pi_R^2} =\frac{1}{4\Omega_R} (|\Omega|^2 + \xi\Omega_R).
\end{equation}
Notice that since we set $k=1$, we are working in a set of coordinates in which $\chi$ and $\Pi$ effectively have the same dimension. To restore dimensions one simply re-inserts factors of $k$. The size of the area in phase space explored by the system is characterised by the \textit{Wigner ellipse} defined by the contour (dropping the $R$ subscript)
\begin{equation}\label{ellipse1}
1 = \frac{2 \left(\Pi+ \Omega _I \chi \right)^2}{\Omega _R + \xi } + 2 \Omega _R \chi ^2,
\end{equation}
This describes an ellipse with a ``tilt" of angle $\alpha$ relative to the $\chi$ axis shown in figure \ref{twoellipse}. The ellipse is aligned with a rotated set of axes $\chi'$ and $\Pi'$, in which its equation is
\begin{equation}\label{ellipse2}
\frac{\chi'^2}{a^2} + \frac{\Pi'^2}{b^2}=1,
\end{equation}
with semi-major and -minor axes $a$ and $b$, respectively. The extreme squeezing of the ellipse amounts gives a one-to-one relation between the values of $\Pi$ and $\chi$ of the form $\Pi \simeq \tan \alpha \chi$. The two sets of coordinates are related by a rotation through an angle $\alpha$:
\begin{equation}
\left(
\begin{array}{c}
\chi'\\
\Pi'
\end{array}
\right)
=
\left(
\begin{array}{cc}
\cos \alpha & - \sin \alpha \\
\sin \alpha & \cos \alpha
\end{array}
\right)
\left(
\begin{array}{c}
\chi\\
\Pi
\end{array}
\right).
\end{equation}
Inserting this expression into (\ref{ellipse2}) and comparing with (\ref{ellipse1}) gives, after a little algebra,
\EQ{
&a^2,b^2 =\\
& \frac{\Omega_R + \xi }{ |\Omega|^2 + \Omega_R \xi+1 \mp \sqrt{(|\Omega|^2 + \Omega_R -1)^2 + 4 \Omega_I^2} },
}
and a rotation angle
\begin{equation}
\tan 2 \alpha = \frac{2 \Omega_R}{1 -|\Omega|^2 - \Omega_R \xi }.
\end{equation}
The area of this ellipse is given by
\begin{equation}
\mathcal{A} = \pi a b = \frac{\pi}{2} \left(1 + \frac{\xi}{\Omega_R}\right)^{1/2}
\end{equation}
From fig.~\ref{twoellipse}, we see that the minor axis is larger in the interacting case due to the system diffusing in phase space. This is reflected by a growth in the Wigner area (fig.~\ref{area}), whose size is determined by the ratio $\xi/\Omega_R$. The growth in the Wigner area is characteristic of open systems and indicates that the state is becoming more ``classical". Indeed, the inverse of the area is roughly the product of the position and momentum correlation lengths and increasing area means that the product of the correlation lengths is becoming smaller than $\hbar(=1)$. In the associated classical stochastic system, the widening of the ellipse is due to diffusion created by the noise in the Langevin equation \eqref{Lang}. The broadening of the Wigner ellipse is also reflected in the momentum correlator $\braket{\Pi_R^2}$ plotted in fig.~\ref{PiPi}. Indeed from eq.~(\ref{correlators}) we see the main contribution to $\braket{\Pi_R^2}$ comes from $\xi$ on super-horizon scales. Thus the broadening of the ellipse corresponds to local effects. By contrast, the length of the ellipse, which characterizes the size of $\braket{\chi^2}$, appears to be roughly the same in the free and interacting cases. The reason for this is that the corrections to $\braket{\chi^2}$ are derived from non-local (memory) contributions, and are therefore much weaker in comparison the local-contributions driving the momentum diffusion and broadening of the ellipse.


\section{Sensitivity of discord to decoherence}\label{discordSec}

Entanglement is a defining feature of quantum systems and is a subtle form of correlation that cannot be mimicked by a classical system. However, it is {\it not\/} an indication of non-local interaction: correlation is {\it not\/} interaction! So entanglement, and its ensuing correlations, are the calling card of a quantum system. Given the notorious sensitivity of entanglement to environmental noise, it is of particular interest to investigate the question of how robust the entanglement created during inflation is to the effects of decoherence. Examining this question will be the objective of the remainder of this paper.
\\
\\
In the present context, by writing the Hamiltonian (\ref{Hamil}) in terms of creation and annihilation operators as
\begin{equation}
H_0 =\frac{k}{2} \left[ \, a_{\Bk } a_{\Bk }^\dagger + a_{-\Bk } a_{-\Bk }^\dagger \right]
+ i \frac{a'}{a}\left[ a_{\Bk }^\dagger a_{-\Bk }^\dagger + a_{\Bk } a_{-\Bk } \right],
\end{equation}
it becomes apparent that the squeezing term involving $a'/a$ leads to the creation of entangled particle pairs with opposite momenta. These field quanta are then separated outside the horizon leading to spacelike field correlations on super-Hubble scales, whose strength is related to the expansion scale $H$. The purpose of the last part of this paper is to examine how quantum correlations are affected by decoherence, which intuitively one expects to render an initially quantum system more classical.

\subsection{Overview of quantum discord}

The entanglement entropy is the most familiar way to capture the strength of \textit{quantum} correlations between two regions $A$ and $B$ of a bipartite system, and provides a simple way of quantifying the degree of entanglement within a system. However, the entanglement entropy only provides such a measure when the state of the total system is pure. The question is how can one measure quantum entanglement between $A$ and $B$ when the total state is not pure. In recent years, a new measure -- \textit{quantum discord} \cite{OZ, HV} -- has been proposed as a means of capturing quantum correlations in just such a situation. Quantum discord exploits the difference in classical and quantum posterior probabilities and the discrepancy, at the quantum level, between two classically equivalent definitions of \textit{mutual information}, defined for classical systems as
\begin{equation}
I(A:B) = H(A) + H(B) - H(A,B),
\end{equation}
where $H$ is the Shannon entropy. Classically, this is equivalent to
\begin{equation}
J(A:B) = H(A) - H(A|B),
\end{equation}
where $H(A|B)$ is the Shannon entropy conditional on $B$ defined by
\begin{equation}
H(A|B) = \sum_{b} P(B=b)H(A |B=b),
\end{equation}
with
\begin{equation}
H(A| B=b) = - \sum_a p(A=a | B=b)\log p\left( A=a | B=b\right).
\end{equation}
The equivalence of these two quantities is ensured by Bayes' theorem
\EQ{
p(A|B=b) = \frac{p(A,B=b)}{p(B=b)}\label{Bayes}.
}
However, at the quantum level, defining a posterior probability is more subtle. Indeed, the act of knowing and measuring the state of $B$ may affect the state of the whole system $\rho_{AB}$! It is this subtlety which lies at the heart of quantum discord. Instead, in the quantum picture, one must first define a notion of conditional entropy, and construct a conditional density matrix. To do this, we must choose a positive operator valued measure (POVM) $\left\{ \Pi^B_n \right\}$ -- this provides a set of measurement outcomes for the system $B$ on which to condition the state of $A$. After a measurement outcome of $\Pi_n$ on $B$ the state is projected according to
\begin{equation}
\rho_{AB} \rightarrow \frac{\Pi^B_n \rho_{AB}\Pi^B_n}{P_n},
\end{equation}
where
\begin{equation}\label{Pn}
P_n = \text{Tr}_{AB} \rho_{AB} \Pi^B_n
\end{equation}
is the probability that the measurement outcome $\Pi^B_n$ occurs and ensures the correct normalisation of $\rho_{A | B=\Pi^B_i}$. Hence the reduced density matrix of $A$, conditional on the measurement outcome $\Pi_n^B$, is
\begin{equation}\label{conditional}
\rho_{A | B=\Pi^B_i} = \text{Tr}_B \frac{\rho_{AB} \Pi^B_n}{P^B_n},
\end{equation}
where we assumed that the POVM is projective in the sense that $\Pi_n^2=\Pi_n$. We can then define a conditional entropy over all possible measurement outcomes in $B$:
\begin{equation}
S(A| B = \lbrace \Pi^B_n\rbrace) = \sum_n P_n S(\rho_{A|B=\Pi^B_n}).
\end{equation}
With this translation to the quantum picture, the natural definition of $J(A : B)_{\Pi^B_n}$ becomes
\begin{equation}
\mathcal{J}(A:B) = S(A) - S(A| B = \lbrace \Pi^B_n\rbrace).
\end{equation}
In fact, this is the Holevo information from $A$ to $B$ \cite{ZZ}. The generalisation of the mutual information is more straightforward and consists simply of replacing Shannon by von Neumann entropy:
\begin{equation}\label{mutual}
\mathcal{I}(A:B) =S(A) + S(B) - S(A,B) .
\end{equation}
The quantum discord is then defined by the difference
\EQ{\label{delta}
\delta(A:B)_{\Pi_n^B} & = \mathcal{I}(A:B) - \mathcal{J}(A:B)_{\Pi^B_n} \\
& = S(B) - S(A,B) + \sum_i P_i S(\rho_{A|B= \Pi_n}).
}
Notice the appearance of the POVM $\lbrace \Pi_n\rbrace$ in the definition, which reminds us how conditional probability is intimately related to measurement at the quantum level.

Note that the discord depends implicitly on the choice of POVM and so one can define {\it the\/} discord by minimizing over the choice of POVM. In general, this minimizing is difficult to perform and below we shall content ourselves with calculating a discord for a particular physically motivated choice of POVM. Our POVM dependent discord is therefore only an upper bound on {\it the\/} discord.

\subsection{Decoherence and Discord in inflation}

\begin{figure}
\begin{center}
\includegraphics{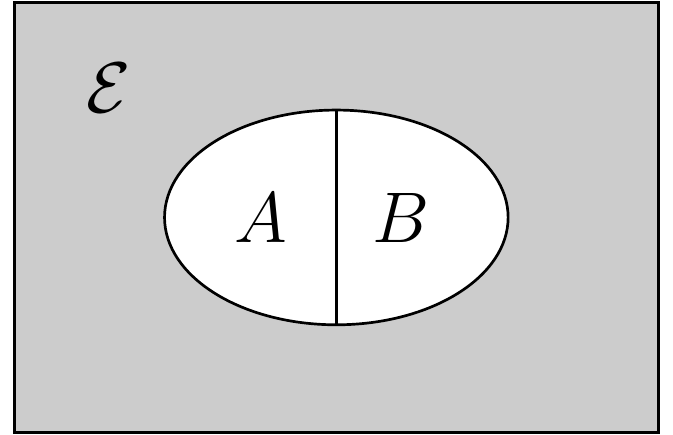}
\caption{\footnotesize The effect of an external environment $\mathcal{E}$ on quantum correlations between $A$ and $B$ in an open bipartite system.}
\label{partition}
\end{center}
\end{figure}

Recently, discord has been applied in the context of cosmological inflation in \cite{Lim} to probe correlations
between primordial fluctuations and environment degrees of freedom and
in \cite{VM} to probe correlations amongst an isolated system of primordial fluctuations.  Our aim is to build on the approach of \cite{VM} and ask what happens to these correlations once the system is exposed to an external decohering environment. This is illustrated by the cartoon in Fig.~\ref{partition}. We shall follow the approach of \cite{VM} and choose the most natural partitioning of the (open) system by splitting the Hilbert space into $+\Bk $ and $-\Bk $ modes. Physically, this corresponds to studying entanglement between field quanta created back-to-back by the de Sitter expansion as they journey outside the horizon and experience decoherence. Since we have an additional environment, we are considering a tri-partition of the overall system: the $+\Bk$ mode, $-\Bk$ mode and environmental field $\psi$.

%

We begin by writing the density matrix in a form which makes this partitioning more transparent, expressing it in terms of the occupation number basis as
\begin{equation}
{\rho}(+ \Bk , - \Bk ) = \sum_{n,m} C_{n, m,|n' , m'} \ket{n_\Bk , m_{-\Bk }} \bra{n'_\Bk , m'_{-\Bk }},
\end{equation}
where $C_{n,m|n',m'} = \bra{n_\Bk , m_{-\Bk }} \rho \ket{n'_\Bk , m'_{-\Bk }}$ are time-dependent coefficients. We must now decide on a choice POVM in order to compute the conditional density matrix (\ref{conditional}). We follow \cite{VM} and choose $\Pi^{-\Bk }_n = \ket{n_{-\Bk }} \bra{n_{-\Bk }}$, which leads to
\EQ{\label{conditional2}
\rho_{A|B=n_{-\Bk }} = \frac{1}{p_n}\text{Tr}_{-\Bk }\left[ \rho \Pi^{-\Bk }_n\right] =\frac{1}{p_n} \sum_{m} p_{nm}\ket{m_{\Bk }} \bra{m_{\Bk }},
}
where
\begin{equation}
p_{nm} = \bra{m_\Bk n_{-\Bk }} \rho \ket{m_\Bk n_{-\Bk }},
\end{equation}
and, in analogy with eq.~(\ref{Pn}),
\EQ{
p_n & = \text{Tr}_{(+\Bk ,-\Bk )} \left[ \rho\, \Pi^{-\Bk }_n \right] \\ & = \sum_{m} \bra{m_\Bk n_{-\Bk }} \rho \ket{m_\Bk n_{-\Bk }}
\equiv \sum_m p_{nm}.
}
After a series of somewhat lengthy calculations described in appendix \ref{DiscordCalc}, we find
\begin{align}\label{pM}
p_ n = 4\Omega_R \frac{\!\! \!\left[1- 2 \Omega_R + |\Omega|^2 + \xi \Omega_R\right]^{m}}{\, \,\left[1+ 2 \Omega_R + |\Omega|^2 + \xi \Omega_R\right]^{m+1}} ,
\end{align}
\begin{widetext}
\EQ{
&p_{nm} =\frac{2(m+n)!}{m! \, n!} \frac{ 2 \Omega_R \, \xi^{m+n} }{\left[ 1 + 2 \Omega_R + |\Omega|^2 + \xi \Omega_R + \xi\right]^{m+n+1}}\\
& \times \, \, \tensor[_2]{F}{_1}\Big[\!-m,\!-n,\!-m\!-\! n,\, \, - \frac{1}{\xi^2}
\left( 1- 2 \Omega_R + |\Omega|^2 + \xi \Omega_R -\xi \right)
\left( 1 + 2 \Omega_R + |\Omega|^2 + \xi \Omega_R + \xi\right) \Big] \label{pNM},
}
\end{widetext}
where we remember that earlier in sec.~\ref{LangevinEQ} we switched to units where $k=1$. To restore dimensionality to (\ref{pM}) and (\ref{pNM}) one simply re-inserts appropriate factors of $k$. Notice there is a non-zero probability for states to have different occupation numbers in the $\pm \Bk $ modes, which cannot happen in the non-interacting case as we begin in the vacuum and particles are created only in pairs with opposite momenta. However, the $\chi \psi^2$ interaction violates $\chi$ number in such a way as to allow differing numbers in the $\pm \Bk $ modes. The state of the reduced density $\rho(\Bk)$ is
\begin{equation}
\rho(\Bk ) = \sum_{n} p_n \ket{n_{\Bk }} \bra{n_{\Bk }}.
\end{equation}
This is a thermal state with temperature
\begin{equation}
\beta_k = - \log\left[ \frac{1- 2 \Omega_R + |\Omega|^2 + \xi \Omega_R }{1+ 2 \Omega_R + |\Omega|^2 + \xi \Omega_R} \right].\label{betak}
\end{equation}
The corresponding entropy $S(\Bk )$ is given by
\begin{equation}\label{Sk}
S(\Bk ) = (1 + \braket{n_k})\log(1 + \braket{n_k}) - \braket{n_k} \log \braket{n_k}
\end{equation}
where
\begin{equation}
\braket{n_\Bk } = \frac{1}{e^{\beta_\Bk }-1 },
\end{equation}
is nothing more than the Bose-Einstein distribution.

\noindent Returning to the conditional entropy (\ref{conditional2}) we have
\begin{align}
S \left(A| B = \{\Pi^{n}_{-\Bk }\}\right)& =-\sum^\infty_{m=0} \frac{p_{mn}}{p_n}\log\Big( \frac{p_{mn}}{p_n}\Big).
\end{align}
After a short calculation we find
\begin{equation}
\mathcal{J} = - 2 \sum^\infty_{n=0} p_n \log p_n + \sum^\infty_{m,n=0} p_{mn}\log p_{m n}.
\end{equation}
On the other hand $S(\Bk)= S(-\Bk) = - \sum_{n} p_n \log p_n$ and $S(\Bk,-\Bk)\equiv S^k_\text{ent}[\chi,\psi]$ is identified with the entanglement entropy of the modes of the field $\chi$ and $\psi$ calculated in sec.~\ref{EntropySec}; hence
\EQ{
{\cal I}=-2\sum_{n=1}^\infty p_n\log p_n-S^k_\text{ent}[\chi,\psi]
}
and then (\ref{mutual}) and (\ref{delta}) give the POVM-dependent discord
\EQ{\label{delta2}
\delta&\equiv \delta(\Bk:-\Bk)_{\Pi_n^{-\Bk}} \\ &= -S^k_\text{ent}[\chi,\psi] - \sum^\infty_{m,n=0} p_{mn}\log p_{m n} .
}

For the non-interacting case we find that $p_{nm}= \delta_{n m} p_n$, which leads to
\begin{equation}\label{deltafree}
\delta_{\text{free}} = - \sum_{n} p^{{\text{free}}}_n \log p_{n}^{{\text{free}}} \equiv S_{\text{free}}(\Bk).
\end{equation}
This is nothing more than the entanglement entropy of $\rho_{\text{free}}(\Bk )$. Indeed, it is a well-known fact (see e.g.~\cite{AD}) that the discord of a bi-partite division of a system in a pure state is equal to the entanglement entropy. On the other hand, when there is interaction with a third system -- in this case the $\psi$ field -- it is not possible to define the entanglement entropy between the $+\Bk$ and $-\Bk$ sub-systems and then one needs quantities like the discord to provide a measure of the quantum correlations between the $+\Bk$ and $-\Bk$ sectors. We should emphasize again that what we have calculated is a discord that depends upon the choice of POVM, albeit that it is a particularly natural choice, and as such it provides an upper bound on the true discord. Nevertheless it does provide a measure of sorts of the quantum correlation between our sub systems.

\begin{figure}[ht]
\begin{center}
\includegraphics{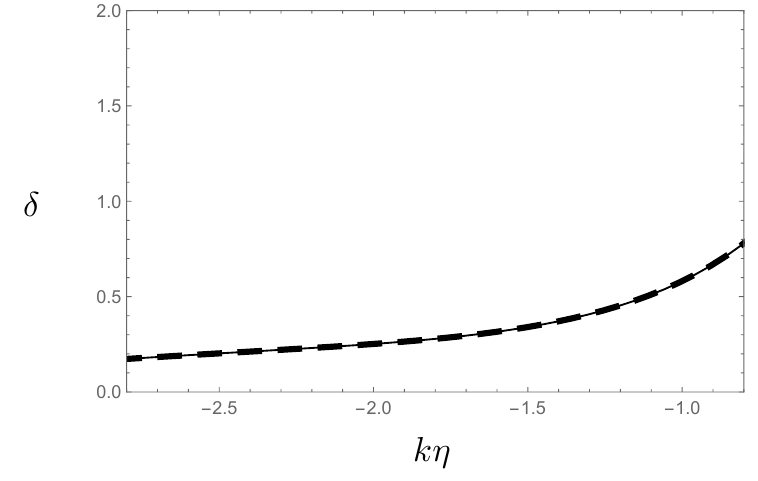}
\caption{\footnotesize POVM dependent quantum discord between the $+\Bk $ and $-\Bk $ modes in the interacting case (solid thin) with $\lambda/H=0.1$ and free case (thick black dashed) $\lambda/H=0$ . }
\label{DiscPlot}
\end{center}
\end{figure}

\begin{figure}[ht]
\begin{center}
\includegraphics{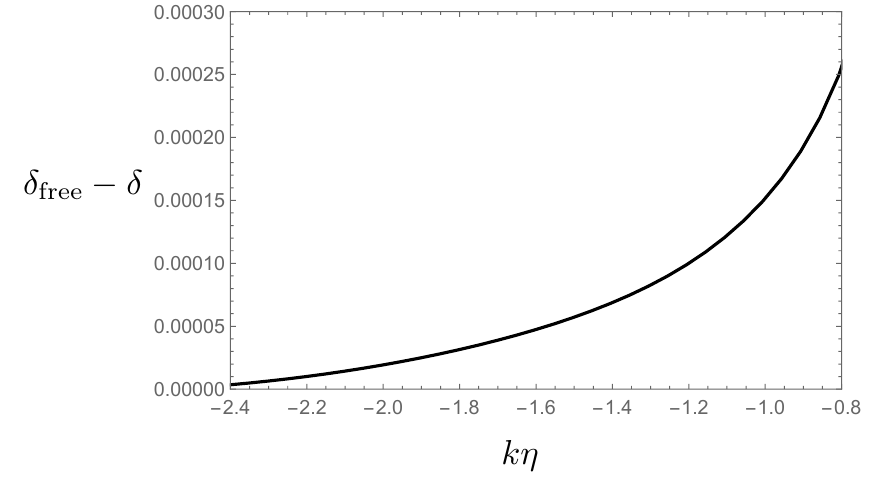}
\end{center}
\caption{\footnotesize The gap (which cannot easily be seen in fig.~\ref{DiscPlot}) between the values of $\delta$ in the interacting case with $\lambda/H=0.1$ and free cases. }
\label{DeltaDiff}
\end{figure}

The discord is plotted in Fig.~\ref{DiscPlot}. Note that the ``free discord" (which is really just the entanglement entropy) dominates the discord in the interacting case, meaning that the $\mathcal{O}\left(\lambda^2/H^2\right)$ corrections to the discord cannot be seen in fig.~\ref{DiscPlot}. The difference between the two is seen more clearly in fig.~\ref{DeltaDiff}. Thus, as expected, we see decoherence \textit{does} have a tendency to weaken discord, though the effect is perturbatively small in our case. One can see this more explicitly by counting powers of $\lambda$. Firstly, we note that since the entropy of a pure state is zero, it follows that
\EQ{
S^k_\text{ent}[\chi,\psi] = \mathcal{O}\left(\lambda^2/H^2\right).
}
Furthermore, we have
\begin{equation}
p_{mn}=
\begin{cases}
p^{\text{free}}_{n} + \mathcal{O}\left(\lambda^2/H^2\right), & \text{if}\ m=n \\
\mathcal{O}\left(\lambda^2/H^2\right), & \text{if}\ m\neq n
\end{cases}
.
\end{equation}
Therefore, from eqs.~(\ref{delta}) and (\ref{deltafree}) we see
\begin{align}
\delta& = \delta_\text{free} + \mathcal{O}\left(\lambda^2/H^2\right).
\end{align}


It has already been noted in other contexts that discord is weakened by the presence of an environment. For example, in \cite{BA}, the effect of various quantum channels acting on spin chains was considered, and it was found that discord in the system was not only weakened by decoherence but actually decayed. This confirms the intuition, that, in a broad sense, decoherence should reduce discord.

For our setup, the discord does not decrease in time, but its growth rate is slowed relative to the free case as can be seen from fig.~\ref{DeltaDiff}.

An interpretation of this might go as follows. In the free case, the discord is just the entanglement entropy between the +$\Bk $ and $-\Bk $ modes. It grows with time as more and more entangled pairs are created, leading the $+\Bk $ and $-\Bk $ sectors to become strongly correlated as time goes on. If one measures $n$ particles in the $-\Bk $ state, this pair-rule ensures that there have to be $n$ particles in state $+\Bk $.

However, in the interacting case, the $\chi \psi^2$ coupling allows the numbers in the $+\Bk $ and $-\Bk $ modes to fluctuate independently beyond pair production due to $\psi \psi \leftrightarrow \chi$ scattering processes. Therefore, if a measurement of the $+\Bk $ mode reveals an occupation number $n$, the state of the $-\Bk $ mode need not contain the same number of particles. This is reflected by the fact that when interactions are switched on, $p_{mn}$ can be non-zero for $n \neq m$. Hence the occupation numbers in the $\pm \Bk $ modes are more weakly correlated in the interacting case due to random environment-induced fluctuations.

Thus two processes -- pair-production and environmental scattering -- are competing to increase and decrease the discord respectively. In the present setup, pair-production wins, leading to an increasing discord albeit at a slower rate in the interacting theory.

One can gain a little insight into this last point with some power counting. The pair production rate of a massless field on dimensional grounds must be of order $H$. Similarly, the scattering rate between a massless field $\chi$ and a conformally coupled field $\psi$ is order $\lambda$. Therefore we expect the correlations between $+\Bk $ and $-\Bk $ sectors to increase if the pair production rate is greater than the scattering rate, i.e., if $1 > \lambda/H$, but this is precisely the perturbativity condition. Hence for $1 >\lambda/H$ we expect the discord to increase. In the $\lambda>H$ case we might expect the opposite, but that would lie beyond the perturbation theory used in our present analysis. One might question how these conclusions would be affected in the case of a different interaction Hamiltonian. We leave such questions for future work.

It is important to realise that, in general, many quantum features -- including entanglement -- are \textit{not} robust to the effects of decoherence. In typical ``laboratory" setups, even the weakest of interactions lead to rapid decoherence and the decay of discord, as shown in \cite{BA}. Therefore, it is far from obvious that quantum correlations should survive the effects of environmental noise. This reveals the importance of the full analysis of decoherence, discord and entanglement presented here. The fact that quantum correlations \textit{do} survive in the present case is due to the continual production of entanglement by cosmological expansion which counteracts the effects of decoherence. It is worth noting that such a mechanism is quite novel to the cosmological scenario.

\section{Discussion}\label{discussion}

In this work we studied the process of decoherence in inflation. By considering an explicit interaction between scalar perturbations and a conformally coupled environment field, we derived the master equation for the reduced density matrix of perturbations first presented in \cite{Boy1}. We have provided a deeper analysis and interpretation of the master equation by explicitly solving for the density matrix with a Gaussian ansatz and Bunch Davies initial conditions. This revealed a scale-dependent nature of decoherence, with the rate at which off-diagonal elements of the density matrix are driven to zero growing in strength on super-Hubble scales. We also found that non-local contributions (i.e. memory effects) from environmental correlators have a negligible effect on the super-horizon strength of decoherence which is dominated by local (white noise) from the environment. By contrast, the Gaussian width receives contributions from non-local memory effects due to long-time environment correlations, which feed into the power spectrum. This confirms from a different perspective the power spectrum calculation of Boyanovksy \cite{Boyanovsky:2015jen}. We also showed that potential positivity-violating effects are suppressed on super-Hubble scales leading to a well behaved density matrix for modes that cross the horizon. This allowed us to construct a set of classical stochastic equations. The resulting Langevin and Fokker-Planck equations provided an emergent classical description of the same dynamics as the quantum master equation on super-Hubble scales. It should be noted that these are due to \textit{genuine} interaction with an environment due to direct couplings between different Fourier modes of the system and environment field and are therefore quite different to the effective stochastic equations encountered in the usual stochastic inflation paradigm \cite{Star1, SY}.

In the remainder of the paper we used these tools to examine the question of the quantum-to-classical transition and the emergence of classicality from an initially pure quantum state. The most natural quantity to consider was the entropy, which grew in time as decoherence took effect. In particular we saw that entropy was determined solely by the ratio $\xi/\Omega_R$, so that the extent to which the state became mixed was controlled by the relative strength of squeezing and decoherence. We also examined the Wigner function and found that the area of the Wigner ellipse $\mathcal{A} = \pi/2 ( 1 + \xi/\Omega_R)^{1/2}$ can be captured by the same ratio $\xi/\Omega_R$, revealing that decoherence has a tendency to increase the size of phase-space explored by the system due to diffusion. Remarkably, however, the power spectrum itself was only affected in a very weak way via non-local memory effects.

Finally, we turned our attention to the question of quantum correlations in the system and how robust they are to environmental noise. We used quantum discord as a measure of the strength of quantum correlations by partitioning the two-mode density matrix into $\pm\Bk $ modes. Physically, this corresponds to the correlations between particle pairs created back-to-back by the de Sitter expansion. It is these correlations which lead to the spacelike field correlations $\braket{\chi_\Bk \chi_{-\Bk } }$ on large scales. By looking at the discord between these modes, we were able to see the effect of decoherence on the strength of quantum correlations between $\pm \Bk $ modes.

For isolated states, pair creation induces strong correlations between the $\pm\Bk $ sectors, but environmental interactions lead to local fluctuations in the numbers in the $\ket{n_\Bk }$ and $\ket{n_{-\Bk }}$ states, reducing the strength of correlations between the two. As a result, decoherence reduces the strength of quantum discord, and in this particular sense, renders the system more classical. However, these corrections were of $\mathcal{O}(\lambda^2/H^2)$, which for our weakly interacting system was perturbatively small and discord remained dominated by the entanglement due to pair-production. This suggests that for weakly interacting systems, at least in the pure de Sitter phase of expansion, pair production dominates over decoherence effects provided the reaction rate with the environment is less than the pair-production rate characterised by $H$. For this reason it is important to reproduce our analysis of discord and decoherence with other interactions.

There are various avenues for developing this work. Firstly, given the time-dependent nature of the interaction Hamiltonian, perturbativity breaks down at late times, meaning that one can only push the analysis used here and in \cite{Boy1} so far outside the Horizon. One resolution to this problem might be to consider other kinds of interactions, e.g.~a marginal coupling of the form $\lambda \int d^4 x \sqrt{-g} \phi \varphi^3 = \lambda \int d^4 x \chi \psi^3$, which does not suffer from the same scaling. This would also provide an interesting way to probe the model-dependence of the growth in decoherence strength observed for the particular interaction considered here.

In this paper we have only considered pure de Sitter expansion, rather than a complete slow-roll analysis which allows inflation to end. Therefore, a more developed analysis of decoherence should include a generic inflationary potential which would allow one to see how decoherence changes as one exists the inflationary era. In particular, one might expect that once the rate of particle pair creation has dropped after inflation, decoherence may persist and wipe out any quantum discord.

Finally, there are many parallels with pair creation in black hole physics. Indeed, both de Sitter and black hole spacetimes possess horizons and exhibit Hawking radiation, which is described by the reduced density matrix after tracing out the remaining member of the pair -- see our eq.~(\ref{betak}). Hence the analysis presented here might provide an interesting springboard for studying decoherence and open quantum-systems in the context of black holes.

\begin{center}
\textbf{Acknowledgments}
\end{center}
JIM would like to thank Gianmassimo Tasinato for first sparking his interest in the question of open systems in inflation, Markus M{\"u}ller for many useful discussions on quantum information, quantum optics and decoherence and Davide De Boni for helpful conversations about Wigner functions, squeezing and decoherence. This work was supported by the STFC grants ST/K502376/1 and ST/L000369/1.

\appendix

\onecolumngrid
\vspace{\columnsep}

\section{Discord calculations}\label{DiscordCalc}

We begin by describing how to compute the matrix elements of the reduced density matrix $\rho(+\Bk )$ in the number basis, defined by
\begin{align}
\bra{n} \rho(+\Bk ) \ket{m} = \sum_{\ell} \bra{ n , \ell} \rho \ket{ m, \ell}
\end{align}
where we have suppressed the $\Bk $ subscripts. Inserting a complete set of position eigenstates on the right hand side gives
\begin{align}
\bra{n} \rho(+\Bk ) \ket{m} = \int d^2x \int d^2y \, \ \rho(y,x) \sum_{\ell} \psi_{m \ell}(x) \psi_{n \ell} ^*(y).
\end{align}
where $\psi_{m l}(x) = \braket{x | m \ell}$ is the position space wavefunction of the number state. This can be written in the usual way, in terms of ladder operators acting on the ground state wavefunction, i.e.
\begin{equation}
\psi_{m, \ell}(x) = \frac{\left( a^\dagger\right)^m}{\sqrt{m!}} \frac{\left( \bar{a}^\dagger\right)^\ell}{\sqrt{\ell!}} \sqrt{\frac{2}{\pi}} e^{- |x|^2}
\end{equation}
where the action of the ladder operators in position space is given by
\begin{equation}
a^\dagger = \frac{x - \bar{\partial}}{\sqrt{2}}, \qquad a = \frac{x + \bar{\partial}}{\sqrt{2}}
\end{equation}
and the bar on the ladder operator denotes complex conjugation, and comes from the ladder operators associated to the $-\Bk $ modes. Notice that since $k$ will drop out of any quantities at the end of the calculation, we have effectively set $k=1$ from the outset. From this it follows that
\begin{equation}
\psi_{m \ell}(x) \psi^*_{n \ell}(y) = \frac{2}{\pi }\frac{1}{\ell!}\frac{1}{\sqrt{n!}\sqrt{m!}} (a^\dagger_x)^m (\bar{a}^\dagger_x)^\ell
(\bar{a}^\dagger_y)^n (a_y^\dagger)^\ell e^{-|x|^2 - |y|^2}.
\end{equation}
Noting that
\begin{equation}
(\bar{a}^\dagger_x a_y^\dagger)^\ell e^{-|x|^2 - |y|^2} = (\sqrt{2} \bar{x} \sqrt{2} y)^\ell e^{-|x|^2 - |y|^2},
\end{equation}
and inserting this expression into the sum over $\ell$ gives
\begin{equation}
\sum_{\ell} \psi_{m \ell}(x) \psi_{n \ell} ^*(y) = \frac{2}{\pi} \frac{1}{\sqrt{m!}\sqrt{n!}} (a^\dagger_x)^m (\bar{a}^\dagger_y)^n e^{-|x|^2 - |y|^2+ 2 \bar{x} y}\ .
\end{equation}
Next one has to consider the action of the remaining derivatives. Using $\bar{a}_y = (\bar{y} - \partial_y)/\sqrt{2}$ gives
\EQ{
\sum_{\ell} \psi_{m \ell}(x) \psi_{n \ell} ^*(y) & = \frac{2}{\pi } \frac{1}{\sqrt{m!}\sqrt{n!}} (a^\dagger_x)^m \left[ \sqrt{2} \left(\bar{y}- \bar{x}\right) \right]^n e^{-|x|^2 - |y|^2+ 2 \bar{x} y}\\
&= \frac{1}{\sqrt{m!}\sqrt{n!}} (a^\dagger_x)^m \left. \frac{d^n}{d \lambda^n} e^{-|x|^2 - |y|^2+ 2 \bar{x} y + \sqrt{2}\lambda(\bar{y} - \bar{x})} \right|_{\lambda =0} .
}
We then act once again with $a_x^\dagger = (x - \bar{\partial}_x) /\sqrt{2}$, giving rise to
\begin{equation}
\sum_{\ell} \psi_{m \ell}(x) \psi_{n \ell} ^*(y) =\frac{2}{\pi }\frac{1}{\sqrt{m!}\sqrt{n!}} \frac{d^n}{d \lambda^n} \left\{ \left[\sqrt{2}(x - y) + \lambda \right]^me^{-|x|^2 - |y|^2+ 2 \bar{x} y + \sqrt{2}\lambda(\bar{y} - \bar{x})} \right\}_{\lambda =0}.
\end{equation}
Now we invoke the Leibniz rule for generalised products and take the $\lambda \rightarrow 0$ limit giving
\begin{equation}
\sum_{\ell} \psi_{m \ell}(x) \psi_{n \ell} ^*(y) = \frac{2}{\pi }\frac{1}{\sqrt{m!}\sqrt{n!}} \sum^n_{k=0}
\left(
\begin{array}{c}
n\\
k
\end{array}
\right)
\frac{m!}{(m-k)!}
\left(\sqrt{2}(x - y) \right)^{m-k} \left( \sqrt{2}\lambda(\bar{y} - \bar{x}\right)^{n-k} e^{-|x|^2 - |y|^2+ 2 \bar{x} y }\ .
\end{equation}
Inserting the matrix elements $\rho(y,x)$ gives
\EQ{
\bra{n} \rho(+\Bk ) \ket{m}& =\frac{2}{\pi } \frac{1}{\sqrt{m!}\sqrt{n!}} \sum^n_{k=0} \frac{n! m!}{(n-k)!(m-k)!} \frac{\sqrt{2}^{m+n-2k}}{k!} \frac{2 \Omega_R}{\pi}\\
& \times \int d^2x \int d^2y \, \, \left(x - y \right)^{m-k} \left(\bar{y} - \bar{x} \right)^{n-k} \exp\Big(- (1 + \Omega)|y|^2 - (1+ \Omega^*) |x|^2 - \frac{\xi}{2}|x -y |^2 + 2 \bar{x} y\Big).
}
The (complex) Gaussian integral is only non-vanishing for $m=n$ which gives
\EQ{
&\bra{n} \rho(+\Bk ) \ket{m} = \delta_{m,n} \frac{4\Omega_R}{\pi^2} \sum^n_{k=0} \frac{n!}{[(n-k)!]^2} \frac{2^{n-k}}{k!} \\
& \times \int d^2x \int d^2y \, \, \left(x \bar{y} + y \bar{x} - |x|^2 - |y|^2 \right)^{n-k} \exp\Big[- (1 + \Omega)|y|^2 - (1+ \Omega^*) |x|^2 - \frac{\xi}{2}|x -y |^2 + 2 \bar{x} y\Big].
}
We can write the Gaussian moments in terms of derivatives, so that
\EQ{
&\bra{n} \rho(+\Bk ) \ket{m} = \delta_{mn}\, \frac{ 4 \Omega_R}{\pi^2 }\sum^n_{k=0} \frac{n!}{[(n-k)!]^2} \frac{2^{n-k}}{k!} \\
& \times \left.\frac{d^{n-k}}{d \, z^{n-k}} \int d^2x \int d^2y \, \exp\Big[- (1 + \Omega+ z)|y|^2 - (1+ \Omega^*+ z) |x|^2 - \frac{\xi}{2}|x -y |^2 + (2+z) \bar{x} y + z \bar{y} x\Big] \right|_{z=0}.
}
Performing the multi-dimensional complex Gaussian integral gives a contribution $\pi^2/\det A$, where $A$ is the covariance matrix appearing in the exponent. Its determinant is given by
\EQ{
\det A =1 + 2 \Omega_R + |\Omega|^2 + \xi \Omega_R + z 2\Omega_R \equiv \alpha + \beta z.
}
From this it follows that
\EQ{
\bra{n} \rho(+\Bk ) \ket{m}& = \delta_{m n} 4 \Omega_R \sum^n_{k=0} \frac{n!}{[(n-k)!]^2} \frac{2^{n-k}}{k!} \left. \frac{d^{n-k}}{d \, z^{n-k}}\left[\alpha + \beta z \right]^{-1} \right|_{z=0} \\
& = \delta_{m,n} 4 \Omega_R \sum^n_{k=0} \frac{n!}{(n-k)!} \frac{2^{n-k}}{k!} (-1)^{n-k} (\beta)^{n-k}\alpha^{-1 - (n-k)} \ .
}
Performing the binomial sum, we get
\begin{equation}
\bra{n} \rho(+\Bk ) \ket{m} = \delta_{m n}4 \Omega_R \frac{\left[ 1 - 2 \Omega_R + |\Omega|^2 + \xi \Omega_R \right]^n}{\left[ 1 + 2 \Omega_R + |\Omega|^2 + \xi \Omega_R\right]^{n+1}} \equiv \delta_{mn} p_n\ .
\end{equation}
Notice this satisfies the correct normalisation $\sum_{n} p_n =1$. The procedure for computing $p_{m n} = \bra{m n} \rho \ket{n m}$ is similar, leading, after a lengthy calculation, to
\EQ{
&p_{nm} =\frac{2(m+n)!}{m! \, n!} \frac{ 2 \Omega_R \, \xi^{m+n} }{\left[ 1 + 2 \Omega_R + |\Omega|^2 + \xi \Omega_R + \xi\right]^{m+n+1}} \\
& \times \, \, \tensor[_2]{F}{_1}\Big[\!-m,\!-n,\!-m\!-\! n,\, \, - \frac{1}{\xi^2}
\left( 1- 2 \Omega_R + |\Omega|^2 + \xi \Omega_R -\xi \right)
\left( 1 + 2 \Omega_R + |\Omega|^2 + \xi \Omega_R + \xi\right) \Big].
}

\vspace{\columnsep}
\twocolumngrid

\end{document}